\documentclass[journal,10pt]{IEEEtran}
\usepackage{amsmath}
\usepackage{soul}
\usepackage{color}
\usepackage{filecontents,lipsum}
\usepackage[noadjust]{cite}
\usepackage{amsthm,amssymb,amsmath,bm}
\usepackage{amsfonts}
\usepackage{epsfig}
\usepackage{epstopdf}
\usepackage{amssymb}
\usepackage{amsmath}
\usepackage{cite}
\usepackage[Algorithm]{algorithm}
\usepackage{color,soul}
\usepackage{multirow}
\usepackage{rotating}
\usepackage{graphicx}
\usepackage{tabularx}
\usepackage{array}
\usepackage{color,soul}
\usepackage{graphicx,dblfloatfix}
\usepackage{blindtext}
\usepackage[dvipsnames]{xcolor}

\newtheorem{remark}{Remark}
\DeclareMathOperator*{\argmax}{arg\,max} 
\usepackage{amsthm,amssymb,amsmath,bm}
\usepackage{graphicx}
\usepackage{fancyhdr}
\usepackage[font={small}]{caption}
\usepackage{subfig}
\usepackage{tabularx}
\usepackage{cite}

\normalsize

\ifCLASSINFOpdf

\else

\fi
\begin{document}
\title{Resource Management and Admission Control for Tactile Internet in Next Generation of RAN}
%
%
%

\author{Narges Gholipoor, Saeedeh Parsaeefard, Mohammad Reza Javan, Nader Mokari, Hamid Saeedi, Hossein Pishro-Nik
\thanks{Narges Gholipoor, Nader Mokari and  Hamid Saeedi are with  Department of Electrical and Computer Engineering, Tarbiat Modares University,	Tehran, Iran}
\thanks{ Saeedeh Parsaeefard is with Iran Telecommunication Research Center, Tehran, Iran}
\thanks{Mohammad Reza Javan is with Department of Electrical and Robotic Engineering, Shahrood University of Technology, Shahrood, Iran}
\thanks{Hossein Pishro-Nik is with Department of Electrical and Computer Engineering, University of Massachusetts Amherst, Massachusetts, United States}
}
%


\maketitle
\begin{abstract}
In this paper, we propose a new queuing model based on admission control (AC) for the Tactile Internet ({TI})  for the cloud radio access network (C-RAN) architecture of the next-generation wireless networks, e.g., 5G, assisted via orthogonal frequency division multiple access (OFDMA) technology. This model includes both the radio remote head (RRH) and baseband processing unit (BBU) queuing delays and reliability for each end to end (E2E) connection between each pair of tactile users.
 In our setup, to minimize the transmit power of users subject to guaranteeing an acceptable delay of users, and fronthaul and access constraints, we formulate a resource allocation (RA) problem. 
    Since the proposed optimization problem is highly non-convex, to solve it efficiently, we employ several transformation techniques such as successive convex approximation (SCA) and the difference of two convex functions (DC). In addition, we propose an AC algorithm to make the problem feasible. In our proposed system model, we dynamically adjust the fronthaul and access links to minimize the transmit power. Simulation results reveal that by dynamic adjustment of the access and fronthaul delays, transmit power can be saved compared to the case of fixed approach per each transmission session. Moreover, the number of rejected users in the network is significantly reduced and more users are accepted.  
\end{abstract}

\vspace{-0.5em}
\IEEEpeerreviewmaketitle
	\section{Introduction}
The Tactile Internet ({TI}) is a new service portfolio of the next generation of wireless networks, e.g., the fifth-generation (5G) wireless networks.
One of the main requirements of {TI} service is ultra-low end-to-end (E2E) delay, e.g., E2E delay should be less than one millisecond \cite{aijaz2017realizing,fettweis2014tactile,she2016ensuring,simsek20165g}.  These requirements cannot be guaranteed via existing wireless networks such as fourth-generation (4G) wireless networks  \cite{simsek20165g}.
However, 5G platform via virtualized and cloud-based architecture can be leveraged to implement the {TI} services  \cite{aijaz2017realizing,fettweis2014tactile}.

 Via the concept of cloud radio access network (C-RAN) in 5G, spectral efficiency (SE), energy efficiency (EE), and cost can be efficiently optimized, where the baseband processing is performed by the baseband units (BBUs) which are connected to remote radio heads (RRHs) through the fronthaul links \cite{park2016joint,7064897}. Specifically, C-RAN reduces energy consumption and increases throughput in dense environment \cite{peng2015system,zhang2016fronthauling}. Therefore, C-RAN is an appropriate architecture for the realization of the {TI} services in dense environment.
In the 5G,  various services with different requirements are realized with a high quality of service (QoS) with the aid of the virtualization techniques. 
Hence, the concept of slicing is introduced in which each slice is defined for a group of users with a set of specific Qos and requirements \cite{7926923,8004168,8115155}.
Slicing provides flexibility to utilize resources which improve SE and EE. However,  in order that the activities of users of one slice do not have detrimental effects on the QoS of the users of other slices, the isolation between slices should be maintained.
There exists various literature for translation to the isolation concept to the proper notation for the networks’ procedures such as dynamic and static methods \cite{7462252,7926923,8004168}. In this paper, we consider a minimum required data rate for each slice to preserve the isolation between slices  \cite{7462252}.

Obviously, for this setup due to the complexity of system architecture, diverse transmission parameters such as power, and different QoS requirements, the problem of resource allocation (RA) is highly essential which has drawn a lot of attention recently \cite{aijaz2016towards,she2016ensuring,she2016uplink,8253477,7945856}. For instance,
in \cite{aijaz2016towards}, a RA problem for the {TI} in the Long-Term Evolution-Advanced (LTE-A) is investigated where the average queuing delay and queuing delay violation in one base station (BS) are optimized. Orthogonal frequency division multiple access (OFDMA) and single carrier frequency division multiple access (SC-FDMA) are considered for downlink (DL) and uplink (UL), respectively. A cross-layer RA  problem for the {TI} is proposed in \cite{she2016ensuring} for single BS where the packet error probability, maximum allowable queuing delay violation probability, and packet dropping probability are jointly optimized with the aim to minimize the total transmit power subject to maximum tolerable queuing delays. In \cite{aijaz2016towards}, queuing delay, packet loss induced by queuing delay violation, packet error, and packet drop caused by channel fading are considered for analyzing the E2E delay of RAN.
In \cite{she2016uplink}, the effect of frequency diversity and spatial diversity on the transmission reliability in UL is studied in the {TI} service where the number of subcarriers, the bandwidth of each subcarrier, and the threshold for each user are optimized for minimizing the total bandwidth to ensure the transmission reliability.
In \cite{8253477}, a multi-cell network based on frequency division multiple access (FDMA) with a fixed delay for backhaul is studied in the {TI} service. Moreover, queuing delay, delay violation probability, and decoding error probability are considered for analyzing the E2E delay of the {TI} service \cite{8253477}.

In the above-mentioned works, a network is considered in which for each user one queue at the BS is assumed. Therefore, by increasing the number of users, a lot of queues are needed at the BS for both UL and DL. However, given that the {TI} is assumed to be implemented in the 5G framework, it is necessary to consider  C-RAN architecture. There exists a set of RRHs in the highly dense network which are connected to BBU center via fronthaul links. Moreover,  previous works consider orthogonal  multiple access techniques instead of NOMA. 
Furthermore, the results in the above works generally ignore the fronthaul delay. However, due to the importance of delay in the {TI}, it is crucial to consider queuing delay in fronthaul as well, otherwise, the resulting allocation of the resources may not practically fulfill the requirement of the {TI}.

To address the mentioned issues, we consider a C-RAN architecture serving a set of tactile users. The contributions of this paper are as follows, many of which have been considered for the first time in the {TI}:

\begin{itemize}
	\item
	We propose a C-RAN scenario in ultra dense environment in 5G platform. This will impose new constraints to the system as far as the number of queues is concerned. For the considered C-RAN architecture, we propose a practical queuing model for sequential queues in the {TI} that can be implemented in realistic networks. 
  Moreover, we consider slicing for the {TI} service in our work. 
  Given that TI services are extremely delay sensitive, there is a possibility that due to sever channel fading, the delay requirements are not met for some tactile users, i.e., the RA problem is not feasible. To tackle this issue and reach an efficient solution, we propose an admission control (AC) where a set of users who has the worst condition to reach a feasible solution is not admitted.
	\item In contrast to \cite{she2016ensuring,she2016uplink,aijaz2016towards} where the fronthaul delay is ignored, we take this delay into consideration. Furthermore,  based on channel state information (CSI), we consider dynamic adjustment of the access and fronthaul queueing delays for each pair of users instead of fixed allowable delay values per each transmission part of our setup and simulation results reveal that it can considerably save the total transmit power.	
\end{itemize}

The rest of this paper is as follows. In Section II, the system model is described. In Section III, we formulate the optimization problem. Numerical results and simulation are presented in Section IV. Finally, Section V concludes the paper. 
\vspace{-1em}
\section{System Model} \label{Systemmodel}
We consider a C-RAN network where all RRHs are connected to the BBU via fronthaul links.
 In this region, there exist several pairs of tactile users where each user aims to send its information to its paired tactile user via the closest RRH through the UL transmission link.
 Then, RRH sends the received data to the BBU via the fronthaul link. The BBU processes all the received data and then sends the data to the corresponding RRH of its paired tactile user. Finally, this RRH transmits the relevant message to the paired tactile user via the DL transmission link.
  Assume each RRH has only one queue for UL transmission and all the data of tactile users is stored in this queue. In addition, we consider only one queue in the BBU to store all received data from RRHs.  
 In DL, we assume that each RRH has a queue for each user for sending data to the paired users.

 \begin{figure}[t] \label{picS}
 	\centering 
 	\includegraphics[width=0.54\textwidth]{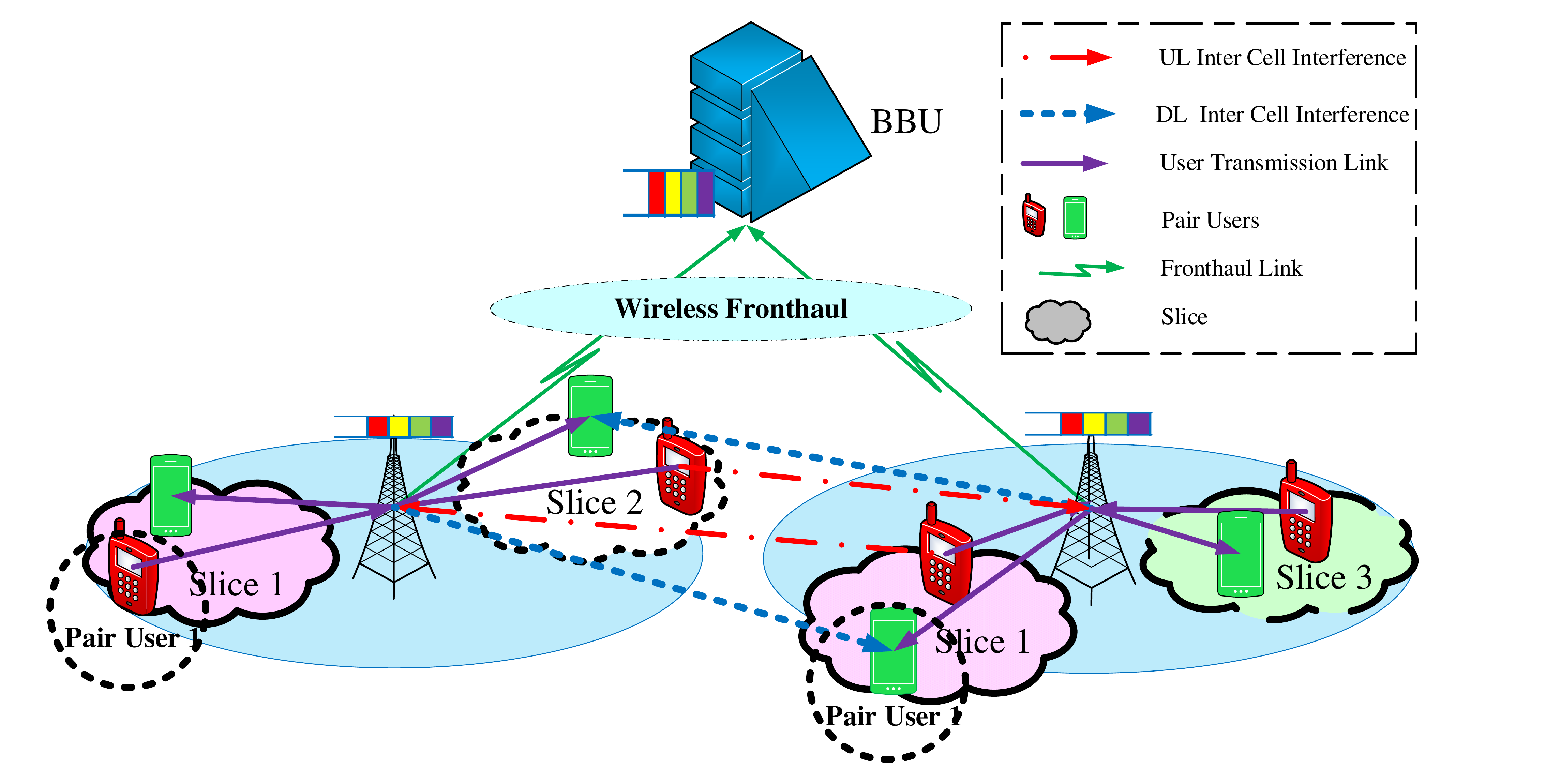}
 	\vspace{-1.5em}
  		\caption{The illustration of the considered network in which three slices with two RRHs are considered. Here as an example, a pair of tactile users is shown by the dotted circles.}
  	\label{pic}
 \end{figure}

As shown in Fig. \ref{pic}, in the considered system model, we have  $\boldsymbol{\mathcal{J}}=\{1,\dots,J\}$ RRHs, $\boldsymbol{\mathcal{S}}=\{1,\dots,S\}$ slices, and  $\boldsymbol{\mathcal{I}}=\{1,...,I\}$ pairs of tactile users.
 Slice $s$ contains $\boldsymbol{\mathcal{I}_s}=\{1,...,I_s\}$ tactile users and the total number of tactile users in our system model is equal to $\boldsymbol{\mathcal{I}}=\bigcup_{s\in \mathcal{S}}{\boldsymbol{\mathcal{I}_s}}$ pairs of users. 
  The terms of access link and fronthaul link often are used to represent the RRH-user connection and RRH-BBU connection, respectively.  In order to overcome the cost of cabling, wireless fronthaul is used instead of fiber fronthaul  \cite{stephen2017joint,park2016joint}. 
  We assume that the fronthaul links are provided via wireless channels in an ultra-dense environment and that there exist two sets of subcarriers $\boldsymbol{\mathcal{K}_1}=\{1,\dots,K_1\}$  and $\boldsymbol{\mathcal{K}_2}=\{1,\dots,K_2\}$ for access and fronthaul links, respectively.
   Moreover, we define $\mathcal{Q}=\{\text{UL},\text{DL}\}$ for simplicity. We consider a two-phase transmission; in the first phase, all tactile users send their data to the corresponding RRH and simultaneously all RRHs send their buffered data to the BBU via fronthaul links.
    In the second phase, all RRHs send data to the corresponding tactile users, and simultaneously, BBU sends the buffered data to all RRHs via fronthaul links.
	These two phases do not perform at the same frequency. Thus, the proposed system model is based on the frequency division duplex (FDD) mode in which each RRH can transmit and receive simultaneously in different frequencies.
	 In order to isolate slices, a minimum required data rate for each slice $s$ must be reserved \cite{7127778,7462252,7324447}. By considering the above definitions, we can now proceed to review the system parameters.
	\vspace{-1em}
\begin{remark}
	 To estimate the CSI for DL links, pilot signals are transmitted via RRHs to all users. Then, each user sends the channel estimation to RRHs via feedback channels. To estimate the CSI for UL, pilot signals are transmitted via users to RRHs, and then, the channel estimations are sent to the users. For the CSI estimation, one of the proposed methods in \cite{coleri2002channel,edfors1996ofdm,li1999pilot} can be utilized.
\end{remark}
\vspace{-2em}
\subsection{Access Links Parameters} \label{Access_Link}
We introduce a binary variable $\tau_{i,k_1}^{s,j,q}$ which is set to 1 if subcarrier $k_1$ is assigned to user $i$ in slice $s$ at RRH $j$, i.e.,
\begin{equation} \nonumber
\begin{split}
&\tau_{i,k_1}^{s,j,q}=\begin{cases}
1, &\text{if subcarrier $k_1$ is assigned to user $i$~in slice $s$}  \\&\text{~at RRH $j$  and $q\in\mathcal{Q}$},\\
0, & \text{otherwise}.
\end{cases}
\end{split}
\end{equation}
Given that we deploy OFDMA in our setup, each subcarrier can be assigned to at most one user. Therefore, we have the following constraint:
\begin{equation} \label{eqo2}
{\text{C1: }}\sum_{s \in \mathcal{S}}\sum_{i \in \mathcal{I}_s}\tau_{i,k_1}^{s,j,q} \le 1, \forall j\in\mathcal{J}, k_1\in\mathcal{K}_1, q\in\mathcal{Q}.\nonumber
\end{equation}
Here, for all $j\in\mathcal{J}$, $k_1\in\mathcal{K}_1$, $s\in\mathcal{S}$, $i\in\mathcal{I}_s$, and $q\in\mathcal{Q}$, the achievable rate for user $i$ on subcarrier $k_1$ at RRH $j$ can be calculated as  \cite{8399832, 8253477}
\begin{equation} \label{eqo3}
\begin{split}
 & r_{i,{k_1}}^{s,j,q} = \frac{{w_{{k_1}}{}}}{{\ln 2}} \bigg [ {\ln (1 + \gamma _{i,{k_1}}^{s,j,q}) - \sqrt {\frac{{V_{i,{k_1}}^{s,j,q}}}{\phi{w_{{k_1}}{}}}} f_Q^{ - 1}(\varepsilon _{i,{k_1}}^{s,j,q})} \bigg ],
\end{split}
\end{equation}
where  
$\gamma^{s,j,q}_{i,k_1}=\frac{p_{i,k_1}^{s,j,q}h_{i,k_1}^{s,j,q}}{\sigma_{i,k_1}^{s,j,q}+ I_{i,k_1}^{s,j,q}}$, in which $p_{i,k_1}^{s,j,q}$, $h_{i,k_1}^{s,j,q}$, and $\sigma_{i,k_1}^{s,j,q}$ represent the transmit power, channel power gain from  RRH $j$ to user $i$  on subcarrier $k_1$ in slice $s$, and noise power, respectively. Also, 
$ I_{i,k_1}^{s,j,q}$  is the inter-cell interference which is equal to 
$ I_{i,k_1}^{s,j,q}=\sum \limits_{\scriptstyle f \in {\cal J}/j} \sum\limits_{v\in\mathcal{S}}\sum\limits_{u\in\mathcal{I}_{v}} \tau_{u,k_1}^{v,f,q}p_{u,k_1}^{v,f,q}h_{i,k_1}^{v,f,q}$. 
Also  
$\phi$, $w_{{k_1}}$, and $f_Q^{ - 1}(.)$ represent time unit, the bandwidth of subcarrier $k_1$, and the inverse of Gaussian-Q function, respectively. Moreover, ${V_{i,{k_1}}^{s,j,q}}$ is defined as  ${V_{i,{k_1}}^{s,j,q}}=1-\frac{1}{(1+\gamma^{s,j,q}_{i,k_1})^2}$. 
Furthermore, in each time unit (short blocklength regime), the total number of transmitted bits of user $i$ at RRH $j$ in slice $s$ over subcarrier $k_1$ is $\Omega=r_{i,{k_1}}^{s,j,q}\phi$.
From \eqref{eqo3}, the error probability $\varepsilon _{i,{k_1}}^{s,j,q}$ can be calculated as follows:
\begin{equation} \label{eqo31} 
\begin{array}{l}
 \varepsilon _{i,{k_1}}^{s,j,q} =  {f_Q} \bigg({\sqrt {\frac{{w_{{k_1}}^{}}}{{V_{i,{k_1}}^{s,j,q}}}} \left[ {\ln (1 + \gamma _{i,{k_1}}^{s,j,q}) - \frac{{\ln 2 \Omega}}{{w_{{k_1}}\phi{}}}} \right]}\bigg),\\ \forall j\in\mathcal{J}, k_1\in\mathcal{K}_1, s\in\mathcal{S}, i\in\mathcal{I}_s, q\in\mathcal{Q}.
\end{array}
\end{equation}
Since the reliability is of fundamental importance for the {TI} services, we consider the following constraint:
\begin{equation} \label{eqo32}
\begin{split}
& \text{C2:~} \varepsilon _{i,{k_1}}^{s,j,q} \le \xi, \forall j\in\mathcal{J}, k_1\in\mathcal{K}_1, s\in\mathcal{S}, i\in\mathcal{I}_s, q\in\mathcal{Q},\nonumber
\end{split}
\end{equation}
where $\xi$ is error probability threshold.
Given that the Q-function does not have a closed-form, we deploy  approximation $\Xi (\gamma _{i,{k_1}}^{s,j,q})\approx {f_Q}(\frac{{\ln (1 + \gamma _{i,{k_1}}^{s,j,q}) - {\rm{ \Omega /(}}{w_{{k_1}}}\phi )}}{{\sqrt {V_{i,{k_1}}^{s,j,q}{{(\ln 2)}^2}/({w_{{k_1}}}\phi )} }})$ as equation \eqref{eqQ-func} 
\begin{figure*}[t]
	\vspace{-1em}
	\begin{equation} \label{eqQ-func}
	\Xi (\gamma _{i,{k_1}}^{s,j,q}) = \left\{ {\begin{array}{*{20}{l}}
		{1,}&{\gamma _{i,{k_1}}^{s,j,q} \le {B_{{k_1}}} - \frac{1}{{2{{\rm A}_{{k_1}}}\sqrt {{w_{{k_1}}}\phi } }}},\\
		{1/2 - {{\rm A}_{{k_1}}}\sqrt {{w_{{k_1}}}\phi } (\gamma _{i,{k_1}}^{s,j,q} - {B_{{k_1}}}),}&{{B_{{k_1}}} - \frac{1}{{2{{\rm A}_{{k_1}}}\sqrt {{w_{{k_1}}}\phi } }} \le \gamma _{i,{k_1}}^{s,j,q} \le {B_{{k_1}}} + \frac{1}{{2{{\rm A}_{{k_1}}}\sqrt {{w_{{k_1}}}\phi } }}},\\
		0,&{{B_{{k_1}}} + \frac{1}{{2{{\rm A}_{{k_1}}}\sqrt {{w_{{k_1}}}\phi } }} \le \gamma _{i,{k_1}}^{s,j,q}},
		\end{array}} \right.
	\end{equation}
	\vspace{-0.5em}
	\hrule
\end{figure*}
where ${{\rm A}_{{k_1}}} = \frac{1}{{2\pi \sqrt {{2^{2(\Omega /({w_{{k_1}}}\phi ))}} - 1} }}$ and ${B_{{k_1}}} = {2^{\Omega /({w_{{k_1}}}\phi )}} - 1$ \cite{8399832, 8253477}.

The total achievable rate in the access links at RRH $j$ is as follows:
\begin{equation} \label{eqo7}
R_{\text{RRH}_j}^{q}=\sum_{s\in\mathcal{S}}\sum_{u\in \mathcal{I}_s}\sum_{k_1\in\mathcal{K}_1}\tau_{u,k_1}^{s,j,q} r_{u,k_1}^{s,j,q}, \forall  j\in\mathcal{J}, q\in\mathcal{Q}.
\end{equation}

Due to the power limitation of each RRH in DL transmission, we have the following constraint:
\begin{equation} \label{eqo71}
{\text{C3: }} \sum_{i\in \mathcal{I}} \sum_{s\in \mathcal{S}} \sum_{k_1\in \mathcal{K}_1}\tau_{i,k_1}^{s,j,\text{DL}} p_{i,k_1}^{s,j,\text{DL}}\le P_{\text{RRH}_j}^{\text{DL}}, \forall  j\in\mathcal{J}.\nonumber
\end{equation}	
Moreover, due to the power limitation of each user, we have
\begin{equation} \label{eqo72}
{\text{C4: }}\sum_{j\in \mathcal{J}} \sum_{s\in \mathcal{S}} \sum_{k_1\in \mathcal{K}_1}\tau_{i,k_1}^{s,j,\text{UL}} p_{i,k_1}^{s,j,\text{UL}}\le P_{\text{USER}_i}^{\text{UL}}, \forall  i\in\mathcal{I}.\nonumber
\end{equation}
\vspace{-1.2em}
\subsection{Fronthaul Links Parameters}
We introduce a binary variable $x_{k_2}^{j,q}$ denoting that subcarrier $k_2$ is assigned to RRH $j$ which is defined by
\begin{equation} \nonumber
\begin{split}
&x_{k_2}^{j,q}=\begin{cases}
1, & \text{if subcarrier $k_2$ is assigned to $\text{RRH}_j$ and $q\in\mathcal{Q}$},\\
0, & \text{Otherwise}.
\end{cases}.
\end{split}
\end{equation}

Assuming that OFDMA is also deployed for the fronthaul links, again each subcarrier can be allocated to at most one RRH, and hence, we have the following constraint:
\begin{equation} \label{eqo9}
{\text{C5: }}\sum_{j \in \mathcal{J}}x_{k_2}^{j,q} \le 1,\forall k_2\in\mathcal{K}_2,q\in\mathcal{Q} .\nonumber
\end{equation}
The achievable rate for each RRH on subcarrier $k_2$ is calculated as follows  \cite{8399832, 8253477}:
\begin{align} \label{eqo10}
\begin{array}{l}
 r_{{k_2}}^{j,q} = \frac{{{w_{{k_2}}}}}{{\ln 2}}\left[ {\ln (1 + \gamma _{{k_2}}^{j,q}) - \sqrt {\frac{{V_{{k_2}}^{j,q}}}{{{\phi  w_{{k_2}}}}}} f_Q^{ - 1}(\varepsilon _{{k_2}}^{j,q})} \right],\\ \forall j \in {\cal J},{k_2} \in {{\cal K}_2},q \in {\cal Q},
 \end{array}\nonumber
\end{align}
where  $\gamma_{k_2}^{j,q}$ is defined as 
$
\gamma_{k_2}^{j,q}=\frac{p_{k_2}^{j,q}h_{k_2}^{j,q}}{\sigma_{k_2}^{j,q}}, \forall j\in\mathcal{J}, k_2\in\mathcal{K}_2,q\in\mathcal{Q}$.  Also $w_{{k_2}}$ is the bandwidth of subcarrier $k_2$ and ${V_{{k_2}}^{j,q}}$ is calculated as  ${V_{{k_2}}^{j,q}}=1-\frac{1}{(1+\gamma^{j,q}_{k_2})^2}$. 
Besides, in each time unit, the total number of transmitted bits is $\tilde\Omega=r_{{k_2}}^{j,q}\phi$.
Similar to the previous subsection (\ref{Access_Link}) and based on \eqref{eqo10}, the error probability $\varepsilon _{{k_2}}^{j,q}$ can be obtained as follows:
	\begin{equation}
	\text{C6:~}\varepsilon _{{k_2}}^{j,q} \le \xi, \forall j \in {\cal J},{k_2} \in {{\cal K}_2},q \in {\cal Q}.
	\end{equation}
Given that the Q-function does not have a closed-form, we deploy  approximation $\tilde \Xi (\gamma _{{k_2}}^{j,q})\approx{f_Q}(\frac{{\ln (1 + \gamma _{{k_2}}^{j,q}) - \Omega {\rm{/}}({w_{{k_2}}}\phi )}}{{\sqrt {V_{{k_2}}^{j,q}{{(\ln 2)}^2}/({w_{{k_2}}}\phi )} }})  $ similar to the previous subsection (\ref{Access_Link}). 
The total achievable rate in the BBU is obtained as follows:
\begin{equation} \label{eqo13}
R_{\text{BBU}}^{q}= \sum_{j\in \mathcal{J}}\sum_{k_2\in\mathcal{K}_2} x_{k_2}^{j,q}r_{k_2}^{j,q}, \forall q\in\mathcal{Q}.
\end{equation}
Due to the power limitation of each RRH in UL transmission, we have
\begin{equation} \label{eqo71}
{\text{C7: }} 
 \sum_{k_2\in \mathcal{K}_2} x_{k_2}^{j, \text{UL}} p_{k_2}^{j,\text{UL}}\le P_{\text{RRH}_j}^{ \text{UL}},\forall  j\in\mathcal{J}.\nonumber
\end{equation}	
Moreover, due to the power limitation of the BBU, we have
\begin{equation} \label{eqo72}
{\text{C8: }}  \sum_{j\in \mathcal{J}} \sum_{k_2\in \mathcal{K}_2} x_{k_2}^{j,\text{DL}} p_{k_2}^{j,\text{DL}}\le P_{\text{BBU}}^{\text{DL}}.\nonumber
\end{equation}

\subsection{Queuing Delay Model}

\begin{figure}[t]
	\centering
	\includegraphics[width=0.50\textwidth]{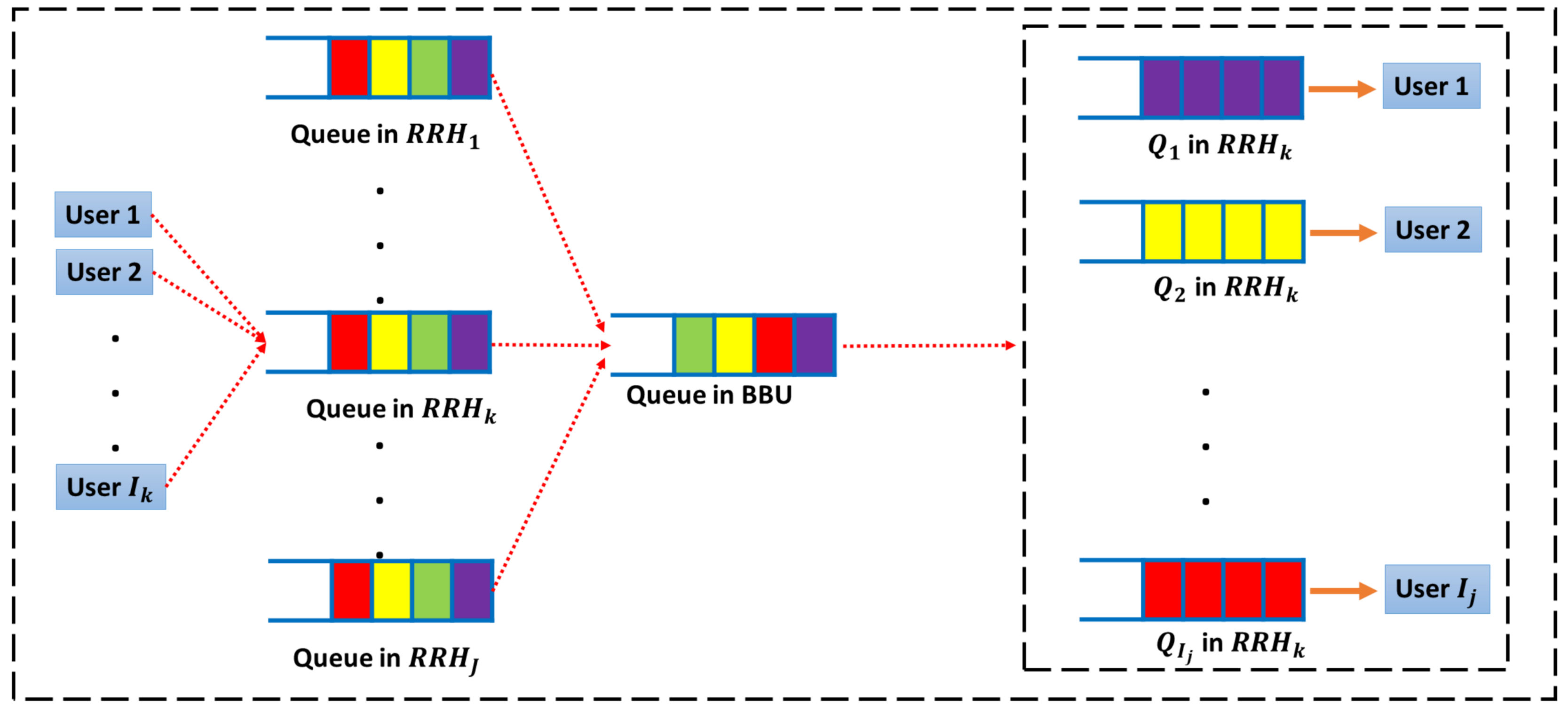}
	\vspace{-1.5em}
	\caption{{Queuing model for our setup where each RRH has only one queue in UL transmission of its all users. RRHs send all data to the BBU Queue. In DL, each RRH has a specific queue for each user.}}
	\label{pic1}
	\vspace{-1.2em}
\end{figure}

The total delay of this architecture consists of three components: delay resulting from UL queues at RRHs, BBU queue, and DL  queues at RRHs, as shown in Fig. \ref{pic1}. Due to delay constraint in the {TI} service, we have
\begin{equation}
{\text{C9: }}D_{\text{max}}^j+D_{\text{max}}^{i,j}+D_{\text{max}}^{BBU}\le D^{\text{max}}_{i,j,s}, \forall i \in \mathcal{I},j\in \mathcal{J}, s\in \mathcal{S},\nonumber
\end{equation}
where $ D_{\text{max}}^j$, $D_{\text{max}}^{i,j}$, $D_{\text{max}}^{BBU}$, and $ D^{\text{max}}_{i,j,s}$ are the delays of UL queues  at RRHs, BBU queue, DL queues at RRH, and the total delay, respectively.
\subsubsection{UL Queuing Delay}
The aggregation of receiving bits from the different transmitters (nodes) can be modeled as a Poisson process \cite{li2007network,she2016ensuring}. The effective bandwidth for a Poisson arrival process in RRH $j$ is defined as \cite{chang1995effective,li2007network,she2016ensuring}
\begin{equation} \nonumber
E_B^{j}(\theta_j)=\lambda_j \frac{(e^{\theta_j}-1)}{\theta_j}, \forall j\in\mathcal{J},
\end{equation}
where $\theta_j$ is the statistical QoS exponent of the $j^{\text{th}}$ RRH. A larger  $\theta_j$ indicates a more stringent QoS and a smaller  $\theta_j$ implies a looser QoS requirement. $\lambda_j$ is the number of bits arrived at RRH $j$ queue defined as 
${\lambda _j} = \sum\limits_{{s} \in {{\cal S}}} {\sum\limits_{u \in {{\cal I}_s}} {\sum\limits_{{k_1} \in {{\cal K}_1}} {r_{u,{k_1}}^{{s,j},\text{UL}}} } },~ \forall j\in\mathcal{J}$.
 The probability of queuing delay violation for RRH $j$ can be approximated as
\begin{equation} \label{deqo17}
\begin{split}
\epsilon_1^{j}= \Pr\{D_j>D^j_{\text{max}}\}=\eta_1 \exp(-\theta_j E_B^{j}(\theta_j) D^j_{\text{max}}),
\end{split}
\end{equation}
for all $j\in\mathcal{J} $ where $D_j$ is the $j^{\text{th}}$ RRH delay, $D^j_{\text{max}}$ is the maximum delay, and $\eta_1$ is the  non-empty buffer probability. Equation \eqref{deqo17} can be simplified to
\begin{equation}\nonumber
\begin{split}
&	\exp ( - {\theta _j}E_B^j({\theta _j}){D_{{\rm{max}}}}) = \exp ( - {\theta _j}{\lambda _j}\frac{{({e^{{\theta _j}}} - 1)}}{{{\theta _j}}}{D^j_{{\rm{max}}}})=\\ &\exp (- {\lambda _j}({e^{{\theta _j}}} - 1){D^j_{{\rm{max}}}}) \le {\delta _1}.
\end{split}
\end{equation}
Therefore, we have	
\begin{equation}
\begin{split}
{\text{C10: }}\sum\limits_{{s} \in {{\cal S}}} {\sum\limits_{u \in {{\cal I}_s}} {\sum\limits_{{k_1} \in {{\cal K}_1}} {r_{u,{k_1}}^{{s,j},\text{UL}}} } }  \ge \frac{{\ln ({1/\delta _1})}}{{( {e^{{\theta _j}}}-1){D^j_{{\rm{max}}}}}}, \forall j \in \mathcal{J}.\nonumber
\end{split}
\end{equation}		
\subsubsection{BBU Queuing Delay}We consider a queue for all RRHs at the BBU for processing data. Therefore, the formulas in the previous section can also be used for this section.
The effective bandwidth for each queue in BBU is $
E_B^{\text{BBU}}(\theta_{\text{BBU}})=\Lambda_{\text{BBU}} \frac{(e^{\theta_{\text{BBU}}}-1)}{\theta_{\text{BBU}}}, 
$
where $\theta_{\text{BBU}}$ is the statistical QoS exponent in the BBU and $\Lambda_{\text{BBU}}$ is the number of bits arrived at  the queue in the BBU which is defined as $
{\Lambda_{\text{BBU}}} ={\sum\limits_{{j} \in {{\cal J}}}}{\sum\limits_{{k_2} \in {{\cal K}_2}} {r_{{k_2}}^{{j},\text{UL}}} }.
$
The probability of queuing delay violation at the BBU can be approximated as
\vspace{-1em}
\begin{equation} \label{deqo19}
\begin{split}
&\epsilon_{\text{BBU}}= \Pr\{D_{\text{BBU}}>D^{\text{BBU}}_{\text{max}}\}=\eta_2 \exp(-\theta_{\text{BBU}}^* E_B^{\text{BBU}}(\theta_{\text{BBU}}) D^{\text{BBU}}_{\text{max}}),
\end{split}
\end{equation}
where $\eta_2$ is the non-empty buffer probability. Equation \eqref{deqo19} can be simplified to
\begin{equation}\nonumber
\begin{split}
&	\exp(-\theta_{\text{BBU}}^* E_B^{\text{BBU}}(\theta_{\text{BBU}}) D^{\text{BBU}}_{\text{max}})= \\&\exp(-\theta_{\text{BBU}}^* \Lambda_{\text{BBU}} \frac{(e^{\theta_{\text{BBU}}}-1)}{\theta_{\text{BBU}}} D^{\text{BBU}}_{\text{max}}) \le {\delta _2}.
\end{split}
\end{equation}
Therefore, we have
\begin{equation}
\begin{split}
{\text{C11: }}{\sum\limits_{{j} \in {{\cal J}}}}{\sum\limits_{{k_2} \in {{\cal K}_2}} {r_{{k_2}}^{{j},\text{UL}}} } \ge \frac{{\ln ({1/\delta _2})}}{{(e^{\theta_{\text{BBU}}}-1){D^{\text{BBU}}_{{\rm{max}}}}}}.\nonumber
\end{split}
\end{equation}
\subsubsection{DL Queuing Delay}
The effective bandwidth for each user in  RRH $j$ is defined as
$
E_B^{i,j}(\theta_i^j)=\lambda_i^j \frac{(e^{\theta_i^j}-1)}{\theta_i^j},~\forall i\in\mathcal{I}, j\in\mathcal{J}, 
$
where $\theta_i^j$ is the statistical QoS exponent of the $i^{\text{th}}$ user in RRH $j$ and  $\lambda_i^j$ is the number of bits arrived at user $i$ queue in RRH $j$ which is defined as $
{\lambda_i^j} = \sum\limits_{{s} \in {{\cal S}}} \sum\limits_{{u} \in {{\cal I}_s}} {\sum\limits_{{k_1} \in {{\cal K}_1}} {r_{u,{k_1}}^{{s,j},\text{DL}}} }, \forall i\in\mathcal{I}, j\in\mathcal{J}$.
The probability of queuing delay violation for user $i$ can be approximated as
\begin{align} \label{ddeqo17}
\epsilon_3^{i,j}= \Pr\{D_i^j>D^{i,j}_{\text{max}}\}=\eta_3 \exp(-\theta_i^j E_B^{i,j}(\theta_i^j) D_{\text{max}}),\nonumber\\ \forall i\in\mathcal{I}, j\in\mathcal{J}, 
\end{align}
where $D_i^j$ is the $i^{\text{th}}$ user delay in RRH $j$ and $\eta_3$ is the  non-empty buffer probability. Equation \eqref{ddeqo17} can simplified to
\begin{equation}\nonumber
\begin{split}
&\exp ( - \theta _i^jE_B^{i,j}(\theta _i^j){D^{i,j}_{{\rm{max}}}}) = \exp ( - \theta _i^j\lambda _i^j\frac{{({e^{\theta _i^j}} - 1)}}{{\theta _i^j}}{D_{{\rm{max}}}}) =\\ &\exp ( - \lambda {_i^j}({e^{\theta _i^j}} - 1){D^{i,j}_{{\rm{max}}}}) \le \delta_3.
\end{split}
\end{equation}
Therefore, we have
\begin{equation}
\begin{split}
{\text{C12: }}\sum\limits_{{k_1} \in {\mathcal{K}_1}} {r_{u,{k_1}}^{s,j,\text{DL}}}  \ge \frac{{\ln (1/\delta_3 )}}{{({e^{\theta _i^j}} - 1){D^{i,j}_{{\rm{max}}}}}}, \forall i\in \mathcal{I}, \forall j \in \mathcal{J}.\nonumber
\end{split}
\end{equation}
In order to avoid bit dropping, the output rate of queues must be greater than the input rate of queues. Therefore, we have two following constraints:
\begin{equation}
{\text{C13: }}  {\sum\limits_{{s} \in {{\cal S}}}  \sum\limits_{u \in {\cal I}_s} {\sum\limits_{j \in {\cal J}} {\sum\limits_{{k_1} \in {{\cal K}_1}} {\tau_{u,k_1}^{s,j,\text{UL}} r_{u,{k_1}}^{{s,j},UL}}}}}\le{\sum\limits_{j \in {\cal J}} {\sum\limits_{{k_2} \in {{\cal K}_2}} {x_{k_2}^{j,\text{UL}}r_{{k_2}}^{{j},{\rm{UL}}}} } },\nonumber
\end{equation}
\begin{equation}
{\text{C14: }}{\sum\limits_{j \in {\cal J}} {\sum\limits_{{k_2} \in {{\cal K}_2}} {x_{k_2}^{j,\text{DL}}r_{{k_2}}^{{j},{\rm{DL}}}} }}\le {\sum\limits_{{s} \in {{\cal S}}}\sum\limits_{u \in {\cal I}_s}{\sum\limits_{j \in {\cal J}} {\sum\limits_{{k_1} \in {{\cal K}_1}} {\tau_{u,k_1}^{s,j,\text{DL}}r_{u,{k_1}}^{{s,j},{\rm{DL}}}} } } }.\nonumber
\end{equation}

\section{Optimization Problem Formulation} 

In this section, our aim is to allocate resources to minimize the overall power consumption in our setup by considering a bounded delay constraint to satisfy the E2E delay requirements. Based on the mentioned constraints C1-C14, the optimization problem can be written as
	\begin{align}
	&\mathop {\min }\limits_{_{\scriptstyle{\bf{P}},{\bf{T}},\hfill\atop
			\scriptstyle{\bf{X}},{\bf{D}}\hfill}} \sum\limits_{j \in {\cal J}} {\sum\limits_{{k_2} \in {{\cal K}_2}} {\sum\limits_{s \in {\cal S}} {\sum\limits_{u \in {{\cal I}_s}} {\sum\limits_{{k_1} \in {{\cal K}_1}} {\sum\limits_{q \in {\cal Q}} {x_{{k_2}}^{j,q}} } } } } } p_{{k_2}}^{j,q} + \tau _{u,{k_1}}^{s,j,q}p_{u,{k_1}}^{s,j,q}\nonumber
	\\\text{s.t.}&:\text{(C1)-(C14)},\label{eqoa}\\ &\text{C15: }\sum\limits_{{k_1} \in {{\cal K}_1}} {\sum\limits_{u \in {\cal I}_s} {\sum\limits_{j \in {\cal J}} {\tau_{u,k_1}^{s,j,q}r_{u,{k_1}}^{{s,j},q}} } }  \ge R^{s, q}_{{\rm{rsv}}},\forall s \in {\cal S},q \in {\cal Q}.~ \nonumber
	\end{align}
The optimization variables in \eqref{eqoa} are subcarrier allocation, power allocation, and delay adjustment for different users in the access and fronthaul as well as in both UL and DL where $\boldsymbol{P}$, $\boldsymbol{T}$, $\boldsymbol{X}$, and $\boldsymbol{D}$ are the transmit power, the access subcarrier allocation, fronthaul subcarrier allocation, and delay vector for users, respectively. The rate constraint C15 is used to isolate the network slices. 

 In problem \eqref{eqoa}, the rate is a non-convex function, which leads to the non-convexity of the problem. In addition, this problem contains both discrete and continuous variables, which makes the problem more challenging. 
 Therefore, we resort to an alternate method to propose an efficient iterative algorithm \cite{4752799,ngo2014joint} with three subproblems, namely, subcarrier allocation subproblem, power allocation subproblem, and delay adjustment subproblem which will be explained in the followings. 
\vspace{-1em}
\section{An Efficient Iterative Algorithm}

 Due to the complex nature of \eqref{eqoa}, and specially having C9-C15, obtaining feasible initial values for problem \eqref{eqoa} is not trivial. Therefore, to find a feasible point for Problem \eqref{eqoa}, we propose to solve the following optimization problem instead of Problem \eqref{eqoa}:

 \begin{align} \label{In_val}
&\begin{array}{l}\mathop {\min }\limits_{\scriptstyle{\bf{P}},{\bf{T}},{\bf{X}},\atop
	\scriptstyle{\bf{D}},{\bf{\alpha }}} \sum\limits_{j \in {\cal J}} {\sum\limits_{{k_2} \in {{\cal K}_2}} {\sum\limits_{s \in {\cal S}} {\sum\limits_{u \in {{\cal I}_s}} {\sum\limits_{{k_1} \in {{\cal K}_1}} {\sum\limits_{q \in {\cal Q}} {\left( {x_{{k_2}}^{j,q}} \right.} } } } } } p_{{k_2}}^{j,q} + \tau _{u,{k_1}}^{s,j,q}p_{u,{k_1}}^{s,j,q}\nonumber\\   + \left. {M\alpha _{u,{k_1}}^{s,j}} \right)\end{array} \nonumber
\\ & \text{s.t.}:\text{C1, C3-C9, C11}, 
\\ &\begin{array}{l}
\text{\~ C2:~} 1 - \exp ( - \frac{{{\gamma ^{{\rm{min}}}}}}{{\bar \gamma^{s,j,q}_{u,k_1} }}) \le \xi+ \alpha ^{s,j}_{u,k_1}, \forall u\in \mathcal{I}, k_1 \in \mathcal{K}_1, \forall j \in \mathcal{J},
\end{array} \nonumber
\\ &\begin{array}{l}{\text{\~ C10: }}\sum\limits_{{s} \in {{\cal S}}} {\sum\limits_{u \in {{\cal I}_s}} {\sum\limits_{{k_1} \in {{\cal K}_1}} {r_{u,{k_1}}^{{s,j},\text{UL}}}+{\alpha_{u,{k_1}}^{{s,j}}} } }  \ge \frac{{\ln ({1/\delta _1})}}{{( {e^{{\theta _j}}}-1){D^j_{{\rm{max}}}}}}, \forall j \in \mathcal{J},\end{array} \nonumber
\\ &\begin{array}{l}{\text{\~ C12: }}\sum\limits_{{k_1} \in {\mathcal{K}_1}} {r_{u,{k_1}}^{s,j,\text{DL}}}+{\alpha_{u,{k_1}}^{{s,j}}}  \ge \frac{{\ln (1/\delta_3 )}}{{({e^{\theta _u^j}} - 1){D^{u,j}_{{\rm{max}}}}}}, \forall u\in \mathcal{I}, \forall j \in \mathcal{J},\end{array} \nonumber 
\\  &\begin{array}{l}{\text{\~ C13: }}  {\sum\limits_{{s} \in {{\cal S}}}  \sum\limits_{u \in {\cal I}_s} {\sum\limits_{j \in {\cal J}} {\sum\limits_{{k_1} \in {{\cal K}_1}} {\tau_{u,k_1}^{s,j,\text{UL}} r_{u,{k_1}}^{{s,j},UL}}}}-{\alpha_{u,{k_1}}^{{s,j}}}}\le\nonumber\\{\sum\limits_{j \in {\cal J}} {\sum\limits_{{k_2} \in {{\cal K}_2}} {x_{k_2}^{j,\text{UL}}r_{{k_2}}^{{j},{\rm{UL}}}} } },\end{array}  \nonumber
\\ &\begin{array}{l}{\text{\~ C14: }}{\sum\limits_{j \in {\cal J}} {\sum\limits_{{k_2} \in {{\cal K}_2}} {x_{k_2}^{j,\text{DL}}r_{{k_2}}^{{j},{\rm{DL}}}} }}\le \nonumber\\{\sum\limits_{{s} \in {{\cal S}}}\sum\limits_{u \in {\cal I}_s}{\sum\limits_{j \in {\cal J}} {\sum\limits_{{k_1} \in {{\cal K}_1}} {\tau_{u,k_1}^{s,j,\text{DL}}r_{u,{k_1}}^{{s,j},{\rm{DL}}}+{\alpha_{u,{k_1}}^{{s,j}}}} } } },\end{array}  \nonumber
\\ &\begin{array}{l}\text{\~ C15:}\sum\limits_{{k_1} \in {{\cal K}_1}} {\sum\limits_{u \in {\cal I}_s} {\sum\limits_{j \in {\cal J}} {\tau_{u,k_1}^{s,j,q} r_{u,{k_1}}^{{s,j},q} +{\alpha_{u,{k_1}}^{{s,j}}} } }}  \ge R^{s, q}_{{\rm{rsv}}},\forall s \in {\cal S}, \end{array}  \nonumber
\end{align}
where $\boldsymbol{\alpha}\ge{0}$ is an elastic variable and if the original problem \eqref{eqoa} is feasible, the optimal value is $\boldsymbol{\alpha}^\star=0$. Furthermore, $M$ is a large coefficient, i.e., $M>>1$. Via this variable, in this section, we propose an AC method to reject users who make the problem infeasible based on a predefined criterion and guarantee the QoS of other users.

Problem \eqref{In_val} is also non-convex and we utilize an iterative algorithm based on the difference of two convex (DC) approximation to transform it into a convex form. 
To solve \eqref{In_val}, we set all the initial values except $\boldsymbol{\alpha}$ to zero and set $\boldsymbol{\alpha}$ to a large value which is a feasible point of \eqref{In_val}. After obtaining the solution of \eqref{In_val}, we check out if the constraints hold or not. If these constraints hold and $\boldsymbol{\alpha}^\star=0$, the derived solution of \eqref{In_val} is an initial value of \eqref{eqoa}; otherwise, we run the AC to reject a user and then repeat this procedure.
\begin{algorithm}[t]
	\caption{Seven-Step Iterative Algorithm}	
	{\textbf{Step 1: Initialization}}
	\begin{itemize}
		\item [] $\mathcal{J} = \{1,...,J\}$, $\mathcal{K}_1 = \{1,...,K_1\}$, $\mathcal{K}_2 = \{1,...,K_2\}$, $\mathcal{I}_s = \{1,...,I_j\}$, $\mathcal{S} = \{1,...,S\}$, $\epsilon_{\text{TH}}=10^{-4}$, $Z_{\text{TH}}=100$
		and $z=0$.
		\item [] Set initial value $p^{(z)}=p^0=0$, $\tau^{(z)}=\tau^0=0$ and  $x^{(z)}=x^0=0$, $\alpha^{(z)}>>0$ .
	\end{itemize} 
	{\textbf{Step 2: Subcarrier Allocation}}
	\begin{itemize}
		\item []	Allocate subcarrier by minimizing the transmit power  and satisfying the problem constraints i.e, \eqref{optimizationtotalsubcarrierllocation11}.
	\end{itemize}
	{\textbf{Step 3: Power Allocation}}
	\begin{itemize}
		\item []Allocate power to each user according to problem (\ref{optimizationtotalpowerllocation1}) and subcarrier allocated in Step 2.
	\end{itemize}
	{\textbf{Step 4: Delay Adjustment}}
	\begin{itemize}
		\item []	Adjust delay of each user  according to problem (\ref{deqoa}).
	\end{itemize}
		{\textbf{Step 5: Finding the Value of $\boldsymbol{\alpha}$}}
		\begin{itemize}
			\item [] Solving problem \eqref{addmission_control}.
		\end{itemize}
		{\textbf{Step 6: Iteration}}
		\begin{itemize}
			\item []	$z=z+1$, Repeat Step 2 to 5 until $||\boldsymbol{P}^{(z)}-\boldsymbol{P}^{(z-1)}||\le\epsilon_{\text{TH}}$ or  $Z_{\text{TH}}<z$.
		\end{itemize}
			{\textbf{Step 7: Admission Control}}
			\begin{itemize}
				\item [] If   $\alpha^{*}\neq0$, reject user $u^{*}$ based on criterion \eqref{AC_Crit}, then return to Step 1.
				\item [] If   $\alpha^{*}=0$, then the algorithm ends.
			\end{itemize}		
		\label{ALG1}
	\end{algorithm}
\begin{figure*}[t]
	\centering
	\includegraphics[width=0.63\textwidth]{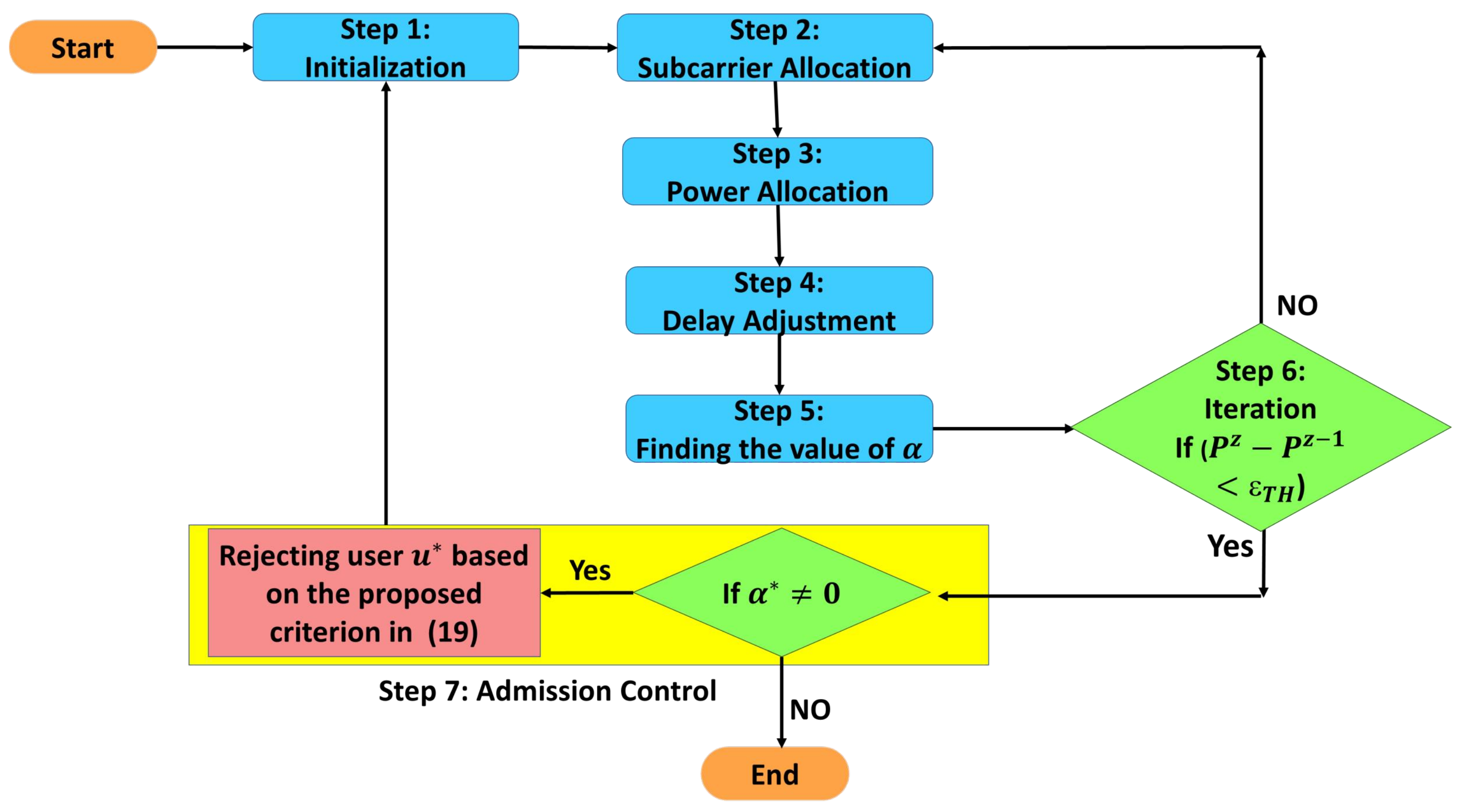}
	\caption{Flowchart of Algorithm.\ref{ALG1}}
	\label{Flowcharttt}
\end{figure*}

As mentioned earlier, to solve \eqref{In_val}, we deploy an iterative algorithm that divides the problem into four subproblems and solve them alternately \cite{4752799,ngo2014joint}.
 This procedure is presented in Algorithm.\ref{ALG1}. Let $z$ be the iteration number and $\boldsymbol{P}^{(0)}$, $\boldsymbol{X}^{(0)}$, and $\boldsymbol{T}^{(0)}$ be the initial values. 
In each iteration, we solve each subproblem by considering the optimization parameters of other subproblems as fixed values derived in the previous steps.
 The iteration stops when the error in Step 6 is less than a predetermined threshold, i.e., $\epsilon_{\text{TH}}$, or the number of iterations exceeds a predetermined value, i.e., $Z_{\text{TH}}$.
  The solution of the last iteration is then declared as the solution of \eqref{In_val}. The flowchart of Algorithm.\ref{ALG1} is shown in Fig. \ref{Flowcharttt}, for which the details are followed.
\\
\textbf{Proposition 1: 
	} The presented iterative algorithm which is described in  Algorithm \ref{ALG1} converges.
\begin{proof} 
	See Appendix A.		
\end{proof}					
\vspace{-1.5em}
\subsection{Subcarrier Allocation Sub-Problem}
With assuming fixed value $\boldsymbol{P}$, $\boldsymbol{\alpha}$  and $\boldsymbol{D}$, the subcarrier allocation subproblem is written as follows:
	\begin{align} \label{optimizationtotalsubcarrierllocation11} 
&\mathop {\min }\limits_{{\bf{T}},{\bf{X}}} \sum\limits_{j \in {\cal J}} {\sum\limits_{{k_2} \in {{\cal K}_2}} {\sum\limits_{s \in {\cal S}} {\sum\limits_{u \in {{\cal I}_s}} {\sum\limits_{{k_1} \in {{\cal K}_1}} {\sum\limits_{q \in {\cal Q}} { {x_{{k_2}}^{j,q}} } } } } } } p_{{k_2}}^{j,q} + \tau _{u,{k_1}}^{s,j,q}p_{u,{k_1}}^{s,j,q} \nonumber
	\\\text{s.t.}:&\text{(C1),(\~ C2),(C3-C8), (\~ C10), (C11), (\~C12-\~C15) }.
	\end{align}
While \eqref{optimizationtotalsubcarrierllocation11} has less computational complexity than \eqref{In_val}, it suffers from  non-convexity due to the interference in the rate functions. In addition, this problem contains discrete variables. We apply time sharing method and relax discrete variables as $x_{k_2}^{j,q}\in[0,1],\forall k_2\in\mathcal{K}_2,\forall j \in \mathcal{J}, \forall q\in\mathcal{Q}$ and $\tau_{u,k_1}^{s,j,q}\in[0,1],\forall u\in\mathcal{I}_s, \forall k_1\in\mathcal{K}_1,\forall j \in \mathcal{J}, \forall q\in\mathcal{Q}$. To solve this problem, we use DC  approximation to transform the problem into a convex form. The subcarrier allocation subproblem is transformed into the following form (See Appendix C):
	\begin{align}\label{4eqo0} 	
	&\begin{array}{l}
	\mathop {\min }\limits_{{\bf{T}},{\bf{X}}} \sum\limits_{j \in {\cal J}} {\sum\limits_{{k_2} \in {{\cal K}_2}} {\sum\limits_{s \in {\cal S}} {\sum\limits_{u \in {{\cal I}_s}} {\sum\limits_{{k_1} \in {{\cal K}_1}} {\sum\limits_{q \in {\cal Q}} { {x_{{k_2}}^{j,q}} } } } } } } p_{{k_2}}^{j,q} + \tau _{u,{k_1}}^{s,j,q}p_{u,{k_1}}^{s,j,q} 	\end{array}\nonumber
	\\\text{s.t.}:&	 \text{(C1),(\~ C2),(C3-C8), (C11) },
\\ & \begin{array}{l}
\text{\={C10}: }\sum\limits_{s \in \mathcal{S}} {\sum\limits_{u \in {\mathcal{I}_s}} {\sum\limits_{{k_1} \in {\mathcal{K}_1}} {f_{{\rm{AC}}}^{{\rm{UL}}}(\tau _{u,{k_1}}^{s,j,{\rm{UL}}}) - g_{{\rm{AC}}}^{{\rm{UL}}}(\tau _{u,{k_1}}^{s,j,{\rm{UL}}})} } }+ \\
\alpha _{u,{k_1}}^{s,j} \ge \frac{{\ln (1/{\delta _1})}}{{({e^{{\theta _j}}} - 1)D_{{\rm{max}}}^j}},\forall j \in J,
\end{array}\text{~~~~~~~~~}\nonumber
	\\ & \begin{array}{l}
	\text{\={C12: } }\sum\limits_{{k_1} \in {\mathcal{K}_1}} {f_{{\rm{AC}}}^{{\rm{DL}}}(\tau _{u,{k_1}}^{s,j,{\rm{DL}}}) - g_{{\rm{AC}}}^{{\rm{DL}}}(\tau _{u,{k_1}}^{s,j,{\rm{DL}}})}  + \alpha _{u,{k_1}}^{s,j} \ge \\
	\frac{{\ln (1/{\delta _3})}}{{({e^{\theta _i^j}} - 1)D_{{\rm{max}}}^{i,j}}},\forall i \in I,\forall j \in J,
	\end{array} \nonumber
\\
 & \begin{array}{l}\text{\={C13: }} {\sum\limits_{{s} \in \mathcal{S}}  \sum\limits_{u \in \mathcal{I}_s} {\sum\limits_{j \in\mathcal{J}} {\sum\limits_{{k_1} \in {\mathcal{K}_1}}}}\tilde f_{\text{AC}}^{\text{UL}}( \tau_{u,{k_1}}^{{s,j},q})-\tilde g_{\text{AC}}^{\text{UL}}( \tau_{u,{k_1}}^{{s,j},q})}+ \\ \alpha_{u,{k_1}}^{{s,j}} +{\sum\limits_{j \in {\cal J}} {\sum\limits_{{k_2} \in {{\cal K}_2}}}} {\log _2}(1 +\frac{{x_{k_2}^{j,\text{UL}}p_{{k_2}}^{{j},{\rm{UL}}}}h_{{k_2}}^{{j},{\rm{UL}}}}{\sigma _{{k_2}}^{{j},{\rm{UL}}}} ) \ge 0, \end{array}\nonumber	
		\\	& \begin{array}{l}
	\text{\={C14: } }  {\sum\limits_{{s} \in {{\cal S}}}\sum\limits_{u \in {\cal I}_s}{\sum\limits_{j \in {\cal J}} {\sum\limits_{{k_1} \in {{\cal K}_1}}}}} f_{\text{AC}}^{\text{DL}}(  \tau _{u,{k_1}}^{{s,j},{\rm{DL}}})- g_{\text{AC}}^{\text{DL}}( \tau _{u,{k_1}}^{{s,j},{\rm{DL}}})\\+\alpha_{u,{k_1}}^{s,j}-{\sum\limits_{j \in {\cal J}} {\sum\limits_{{k_2} \in {{\cal K}_2}}}} g_{\text{FH}}^{\text{DL}} (x_{{k_2}}^{{j},{\rm{DL}}})\ge 0,	\end{array} \nonumber	
		\end{align}
		\begin{align} & \begin{array}{l} \text{\={C15: }}\sum\limits_{s \in S} {\sum\limits_{u \in {I_s}} {\sum\limits_{{k_1} \in {K_1}} {f_{{\rm{AC}}}^{\rm{q}}(\tau _{u,{k_1}}^{s,j,{\rm{q}}}) - g_{{\rm{AC}}}^{\rm{q}}(\tau _{u,{k_1}}^{s,j,{\rm{q}}})} } + \alpha _{u,{k_1}}^{s,j}} \\  \ge R_{{\rm{rsv}}}^{s,q},\forall s \in S,q \in Q.\end{array} \nonumber
	\end{align}
The above problem is a convex problem and can be solved with the CVX toolbox in Matlab \cite{grant2008cvx,boyd2004convex}.
\textbf{Proposition 3
}: The proposed iterative algorithm based on the SCA method for subcarrier allocation subproblem converges.
\begin{proof}
	See Appendix B by considering fixed value for power ($\bf{P}$).
\end{proof}
\vspace{-1em}
\subsection{Power Allocation Sub-Problem}
For the fixed value of $\boldsymbol{T}$, $\boldsymbol{X}$,  $\boldsymbol{\alpha}$ and $\boldsymbol{D}$ the power allocation subproblem is obtained as follows
	\begin{align}
	\label{optimizationtotalpowerllocation1}
		&\mathop {\min }\limits_{{\bf{P}}} \sum\limits_{j \in {\cal J}} {\sum\limits_{{k_2} \in {{\cal K}_2}} {\sum\limits_{s \in {\cal S}} {\sum\limits_{u \in {{\cal I}_s}} {\sum\limits_{{k_1} \in {{\cal K}_1}} {\sum\limits_{q \in {\cal Q}} { {x_{{k_2}}^{j,q}} } } } } } } p_{{k_2}}^{j,q} + \tau _{u,{k_1}}^{s,j,q}p_{u,{k_1}}^{s,j,q} \nonumber
	\\\text{s.t.}:&\text{(\~ C2),(C3-C4),(C6-C8) (\~ C10), (C11), (\~C12-\~C15) }, 
	\end{align}
Similar to the subcarrier allocation subproblem,  in problem \eqref{optimizationtotalpowerllocation1}, the rate is a non-convex function, which leads to the non-convexity of the problem. Therefore, it is necessary to approximate \eqref{optimizationtotalpowerllocation1} with a convex problem. To solve this problem, we use the DC approximation to transform the problem into a convex form.  Therefore, the power allocation subproblem is transformed as follows (See Appendix C):  
	\begin{align}\label{4eqo0p} 		&\mathop {\min }\limits_{{\bf{P}}} \sum\limits_{j \in {\cal J}} {\sum\limits_{{k_2} \in {{\cal K}_2}} {\sum\limits_{s \in {\cal S}} {\sum\limits_{u \in {{\cal I}_s}} {\sum\limits_{{k_1} \in {{\cal K}_1}} {\sum\limits_{q \in {\cal Q}} { {x_{{k_2}}^{j,q}} } } } } } } p_{{k_2}}^{j,q} + \tau _{u,{k_1}}^{s,j,q}p_{u,{k_1}}^{s,j,q} \nonumber
	\\\text{s.t.}:&	 \text{(\~ C2),(C3-C4),(C6-C8) (C11) },
\\& \begin{array}{l} \text{\~{C10: } }  \sum\limits_{s \in \mathcal{S}} {\sum\limits_{u \in {\mathcal{I}_s}} {\sum\limits_{{k_1} \in {\mathcal{K}_1}} {f_{{\rm{AC}}}^{{\rm{UL}}}(p _{u,{k_1}}^{s,j,{\rm{UL}}}) - g_{{\rm{AC}}}^{{\rm{UL}}}(p _{u,{k_1}}^{s,j,{\rm{UL}}})}}}+\\   \alpha _{u,{k_1}}^{s,j}  \ge \frac{{\ln (1/{\delta _1})}}{{({e^{{\theta _j}}} - 1)D_{{\rm{max}}}^j}},\forall j \in {\cal J}, \end{array}\nonumber
	\\ &\begin{array}{l}  \text{\~{C12: } }\sum\limits_{{k_1} \in {\mathcal{K}_1}} {f_{{\rm{AC}}}^{{\rm{DL}}}(p _{u,{k_1}}^{s,j,{\rm{DL}}}) - g_{{\rm{AC}}}^{{\rm{DL}}}(p _{u,{k_1}}^{s,j,{\rm{DL}}})}  + \alpha _{u,{k_1}}^{s,j} \ge\\ \frac{{\ln (1/{\delta _3})}}{{({e^{\theta _i^j}} - 1)D_{{\rm{max}}}^{i,j}}},\forall i \in {\cal I},\forall j \in {\cal J},\end{array}\nonumber  
	\\ & \begin{array}{l} \text{\~{C13: } } {\sum\limits_{{s} \in {{\cal S}}}  \sum\limits_{u \in {\cal I}_s} {\sum\limits_{j \in {\cal J}} {\sum\limits_{{k_1} \in {{\cal K}_1}}}}}\tilde f_{\text{AC}}^{\text{UL}}( p_{u,{k_1}}^{{s,j},q})-\tilde g_{\text{AC}}^{\text{UL}}( p_{u,{k_1}}^{{s,j},q})+\\ \alpha_{u,{k_1}}^{{s,j}}+{\sum\limits_{j \in {\cal J}} {\sum\limits_{{k_2} \in {{\cal K}_2}}}} {\log _2}(1 +\frac{{x_{k_2}^{j,\text{UL}}p_{{k_2}}^{{j},{\rm{UL}}}}h_{{k_2}}^{{j},{\rm{UL}}}}{\sigma _{{k_2}}^{{j},{\rm{UL}}}} ) \ge 0, \end{array}\nonumber 
\\& \begin{array}{l} \text{\~{C14: } } {\sum\limits_{{s} \in {{\cal S}}}\sum\limits_{u \in {\cal I}_s}{\sum\limits_{j \in {\cal J}} {\sum\limits_{{k_1} \in {{\cal K}_1}}}}} f_{\text{AC}}^{\text{DL}}( p_{u,{k_1}}^{{s,j},{\rm{DL}}})- g_{\text{AC}}^{\text{DL}}(p _{u,{k_1}}^{{s,j},{\rm{DL}}})+\\\alpha_{u,{k_1}}^{{s,j}}-{\sum\limits_{j \in {\cal J}} {\sum\limits_{{k_2} \in {{\cal K}_2}}}} g_{\text{FH}}^{\text{DL}}(p_{{k_2}}^{{j},{\rm{DL}}})\ge 0, \end{array}\nonumber 
 \\ &\begin{array}{l} \text{\~{C15: } }\sum\limits_{s \in \mathcal{S}} {\sum\limits_{u \in {\mathcal{I}_s}} {\sum\limits_{{k_1} \in {\mathcal{K}_1}} {f_{{\rm{AC}}}^{\rm{q}}(p _{u,{k_1}}^{s,j,{\rm{q}}}) - g_{{\rm{AC}}}^{\rm{q}}(p _{u,{k_1}}^{s,j,{\rm{q}}})}  + \alpha _{u,{k_1}}^{s,j}} }\\  \ge R_{{\rm{rsv}}}^{s,q},\forall s \in S,q \in Q. \end{array}\nonumber 
	\end{align}
Similar to the subcarrier allocation subproblem, the above problem is a convex problem  and can be solved with the CVX toolbox in Matlab \cite{grant2008cvx,boyd2004convex}.
\\
\textbf{Proposition 3
}: The proposed iterative algorithm based on the SCA method for power allocation subproblem converges.
\begin{proof}
	See Appendix B by considering fixed values for subcarrier allocation parameters i.e, ($\bf{T}$) and ($\bf{X}$).
\end{proof} 
\vspace{-1.5em}
\subsection{Delay Adjustment Sub-Problem}
With assuming fixed values of $\boldsymbol{P}$, $\boldsymbol{T}$, $\boldsymbol{\alpha}$ and $\boldsymbol{X}$, the delay adjustment subproblem is obtained as
\vspace{-0.5em}
	\begin{align}\label{deqoa}
	&\text{find}  \,\,\,\,\, {\boldsymbol{D}} 
	\\		\text{s.t.}:& \text{(C9), (\~C10), (C11),  (\~C12)}. \nonumber
	\end{align}
The delay adjustment subproblem can be solved by the linear programming (LP) of any optimization toolbox.
\vspace{-1.5em}
\subsection{Finding the Value of $\boldsymbol{\alpha}$ }
To find the optimal value for $\boldsymbol{\alpha}$, we have to solve the following subproblem
	\begin{align}
	\label{addmission_control}
	&\min_{\boldsymbol{\alpha}} \sum_{j\in \mathcal{J}}\sum_{s\in\mathcal{S}}\sum_{u\in \mathcal{I}_s}\sum_{k_1\in\mathcal{K}_1}\alpha_{u,k_1}^{s,j}
	\\\text{s.t.}:&\text{ (\~ C10), (\~C12-\~C15)}. \nonumber
	\end{align}
	Similar to the delay adjustment problem, this problem can be solved by the linear programming (LP) of any optimization toolbox.
	\vspace{-1em} 	
{\subsection{Admission Control (AC) }\label{Initial}
		In this paper, we deploy an AC method to make the problem feasible.
		As shown in Fig .\ref{Flowcharttt}, in this method, when the problem is infeasible, the user that has the most impact on infeasibility is recognized based on a newly defined criterion and then, this user will be rejected. Then the problem is resolved for other remaining users. This process continues until the problem becomes feasible. 
It's worth noting that if $\alpha^{*}=0$ (step 5 of Fig .\ref{Flowcharttt}), it means that all the constraints are satisfied and there is no need to reject any user. However, for $\alpha^{*}\neq0$, all constraints of problem \eqref{eqoa} are not satisfied. Therefore, the algorithm {rejects user $u^*$ which enforces} the highest level of infeasibility to the problem based on the following criterion (returns to Step 1 of Fig .\ref{Flowcharttt}):
\begin{equation}\label{AC_Crit}
u^*=\argmax_{u\in \mathcal{I}}  \alpha_{u,{k_1}}^{s,j}
\end{equation}
}
\vspace{-3em}
\subsection{Computational Complexity}
 The number of required iterations for the DC approximation is $ \frac{\log C\ {t^0\varrho}}{\xi}$, where $0\le\varrho\le \infty$ is the stopping criterion for the interior point method (IPM), $\xi$ is used to update accuracy of the IPM, $t^0$ is initial point for approximating the accuracy of IPM, and $C$ is the total number of constraints. For the subcarrier allocation subproblem, the total number of constraints is denote by $C_\text{Sub}$ which is $C_\text{Sub}=2J{K_1} + 2J{K_1}I + 2{K_2} + 2J{K_2} + 3J + I + JI + 2S + 4$ \cite{ngo2014joint}. Similarly, for the power allocation subproblem, the total number of constraints is $C_\text{Pow}=2J{K_1}I + 3J + I + JI + 2J{K_2} + 2S + 4$. For the delay adjustment subproblem, the total number of constraints is $C_\text{Delay}=J+2JI+1$. In AC subproblem, the total number of constraints is $C_\text{AC}=J+IJ+2S+3$. For instance, in subcarrier allocation subproblem and power allocation subproblem, the number of RRHs and the number of users have a significant effect on the computational complexity. Moreover, the subcarrier allocation subproblem has more computational complexity than other subproblems. In contrast, the AC subproblem has lower complexity than the others.
\vspace{-1em}
\section{Simulation and Results}
In this section, the simulation results are presented to evaluate the performance of the proposed system model. To simulate dense urban area, we consider a BBU is at the center of the coverage area whose distance is $1$~Km from a set of RRHs. The coverage area is considered $10$ square Kilometers. Moreover, we consider a Rayleigh fading wireless channel in which the subcarrier gains are independent. Channel power gains for the access links are set as $h_{i,k_1}^{s,j,q}=\Omega_{i,k_1}^{s,j,q}{d_{i}^{j,q}}^{-\alpha}$ where $d_{i}^{j,q}$ is the distance between user $i$ and RRH $j$, $\Omega_{i,k_1}^{s,j,q}$ is a random variable which is generated by Rayleigh distribution, and $\alpha=3$ is the path-loss exponent. Channel power gains for fronthaul links are set to $h_{k_2}^{j,q}=\Omega_{k_2}^{j,q}{d_{j,q}}^{-\beta}$ where similar to access links, $d_{j,q}$ is the distance between RRH $j$ and BBU, $\Omega_{k_2}^{j,q}$ is a random variable generated according to the Rayleigh distribution, and $\beta=3$ is the path-loss exponent. The power spectral density (PSD) of the received Gaussian noise is set to $-174$~dBm/Hz. 
At each RRH, we set $P_{\text{RRH}_j}^{\text{DL}}=43$~dBm $\forall j\in\mathcal{J}$ and $P_{\text{RRH}_j}^{\text{UL}}=43$~dBm $\forall j\in\mathcal{J}$. For the BBU, we set $P_{\text{BBU}}^{\text{DL}}=46$~dBm, and for each user, we set $P_{\text{USER}_i}^{\text{UL}}=23$~dBm \cite{8399832, 8253477}. The frequency bandwidth of wireless access and fronthaul links are $W_\text{AC}=100$~MHz and $W_\text{FH}=100$~MHz, respectively. Moreover, the bandwidth of each subcarrier is $W_S=2$~MHz \cite{8399832,8253477}. The QoS exponent is $\theta=10$ \cite{ngo2014joint}. Furthermore, we assume packet size is equal to $20$~bytes.  In this section, we use Monte-Carlo method for simulation where the optimization problem is solved for 1000 channel realizations  and the total transmit power is the average value over all derived solutions.
\vspace{-1em}
\subsection{Effects of Network Parameters}
 Here, we investigate the effects of network parameters on the performance of our setup based on OFDMA assisted C-RAN. In this regard, we introduce the percentage of service acceptance ratio (SAR) criterion to evaluate the performance of the proposed system model. If the number of users who request the service is $I$, and the number of admitted users is $\tilde I$, the percentage of SAR criterion is equal to $ 100 \times \dfrac{\tilde I}{I}\%$. In this section, we also evaluate this criterion for the network performance analysis in addition to the total transmit power.

Unless otherwise stated, we consider 3 RRHs, 2 slices, $1$~ms E2E delay and packet error rate (PER) of about $10^{-7}$.
\begin{figure*}[t]
		\vspace{-2em}
	\centering
	\subfloat[Total transmit power vs. total number of users per cell]{\includegraphics[width=0.45\textwidth]{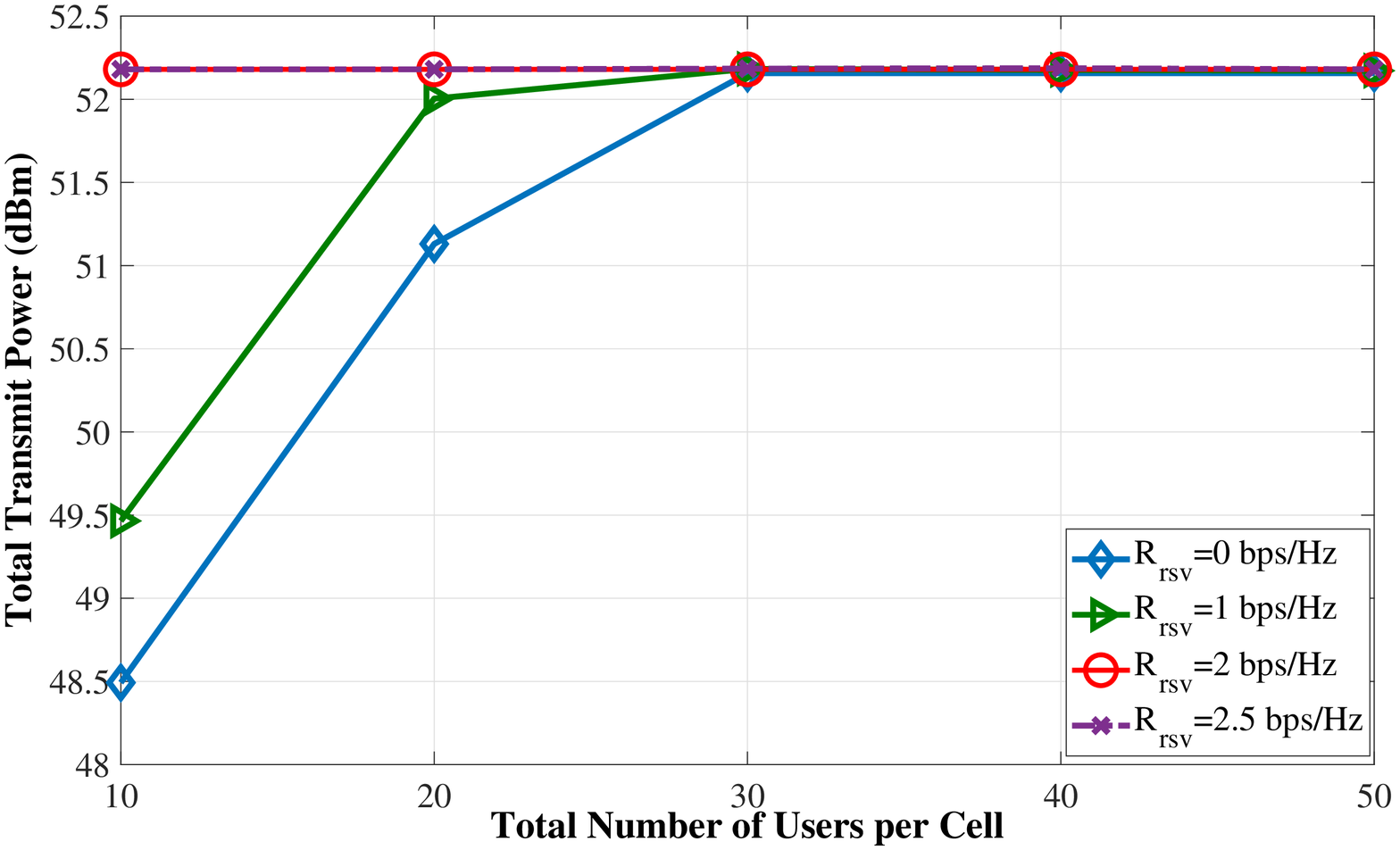}\label{Sim5}}
	\subfloat[Service acceptance ratio vs. total number of users per cell]{\includegraphics[width=0.45\textwidth]{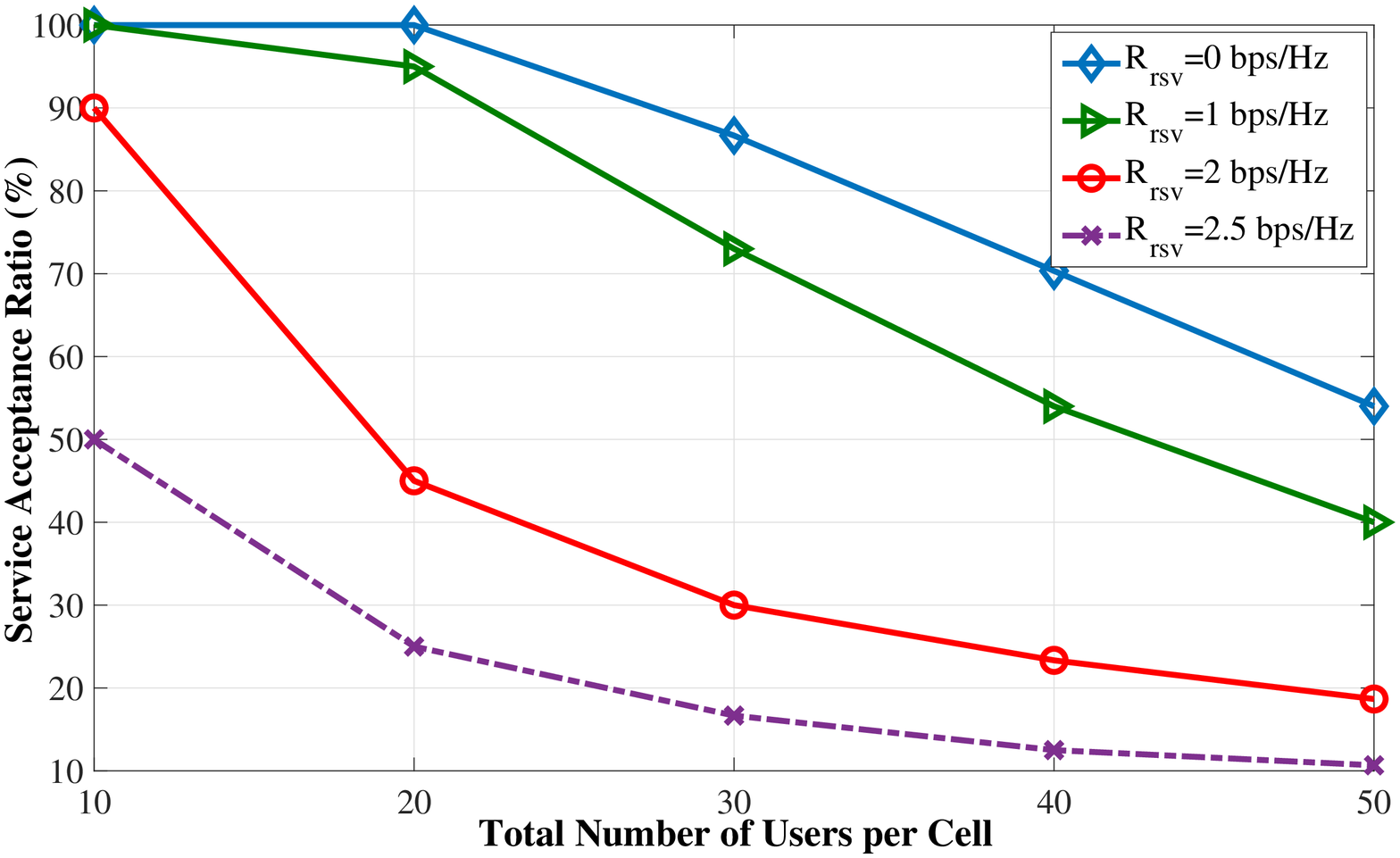}\label{Sim6}}	
	\vspace{-0.5em}
	\caption{System performance versus total number of users}
	\label{Sim56}
\end{figure*}	
Fig. \ref{Sim56}(a) shows the total transmit power versus the total number of users per cell for different reservation rates $R_{\text{rsv}}$.
  As expected, the total transmit power increases by increasing the number of users per cell.
  It stems from the fact that each user has its own QoS determined by the corresponding delay requirement.
    Moreover, the total transmit power increases by increasing the value of the reservation rate  $R_{\text{rsv}}$. Fig. \ref{Sim56}(b) shows the {SAR} versus the total number of users per cell for different reservation rates $R_{\text{rsv}}$. 
    As expected, when the number of users is increased, the amount of {SAR} reduces. On the other hand, increasing the reservation rate leads to degradation in the {SAR}. 
For low reservation rates, e.g., $R_{\text{rsv}}\in \{0, 1\}$~bps/Hz, when a few number of users exist in the network, all the users' requested services are accepted. 
    In contrast, for high reservation rates, e.g., $R_{\text{rsv}}\in \{2, 2.5\}$~bps/Hz, the reason for not accepting all requested services is the lack of satisfying the reservation rate constraint. Thus, by increasing the reservation rate $R_{\text{rsv}}$ the {SAR} is decreased.
\begin{figure*}[t]
	\centering
	\subfloat[Total transmit power vs. Packet error rate (PER)]{\includegraphics[width=0.45\textwidth]{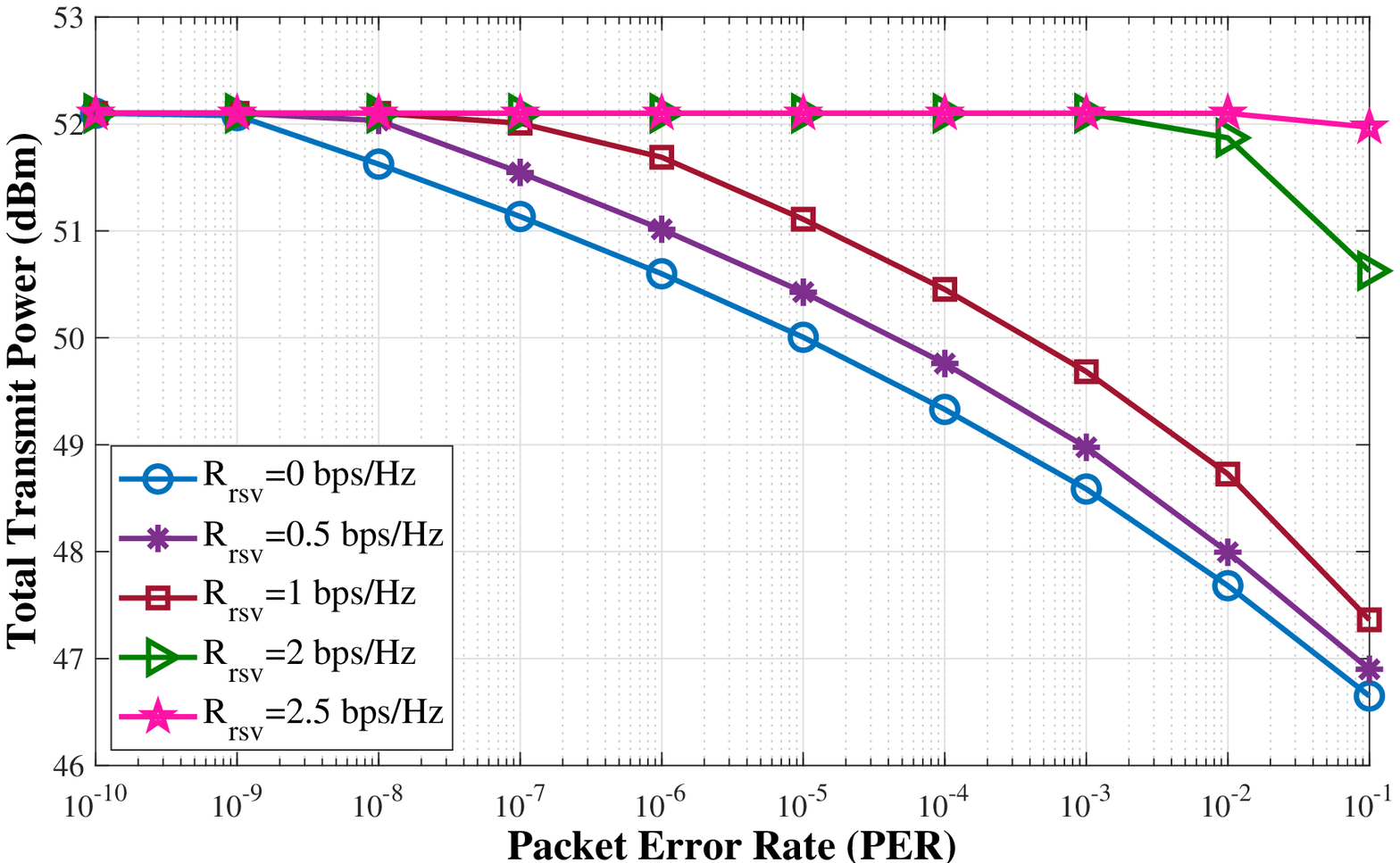}\label{Sim7}}
	\subfloat[Service acceptance ratio vs. Packet error rate (PER)]{\includegraphics[width=0.45\textwidth]{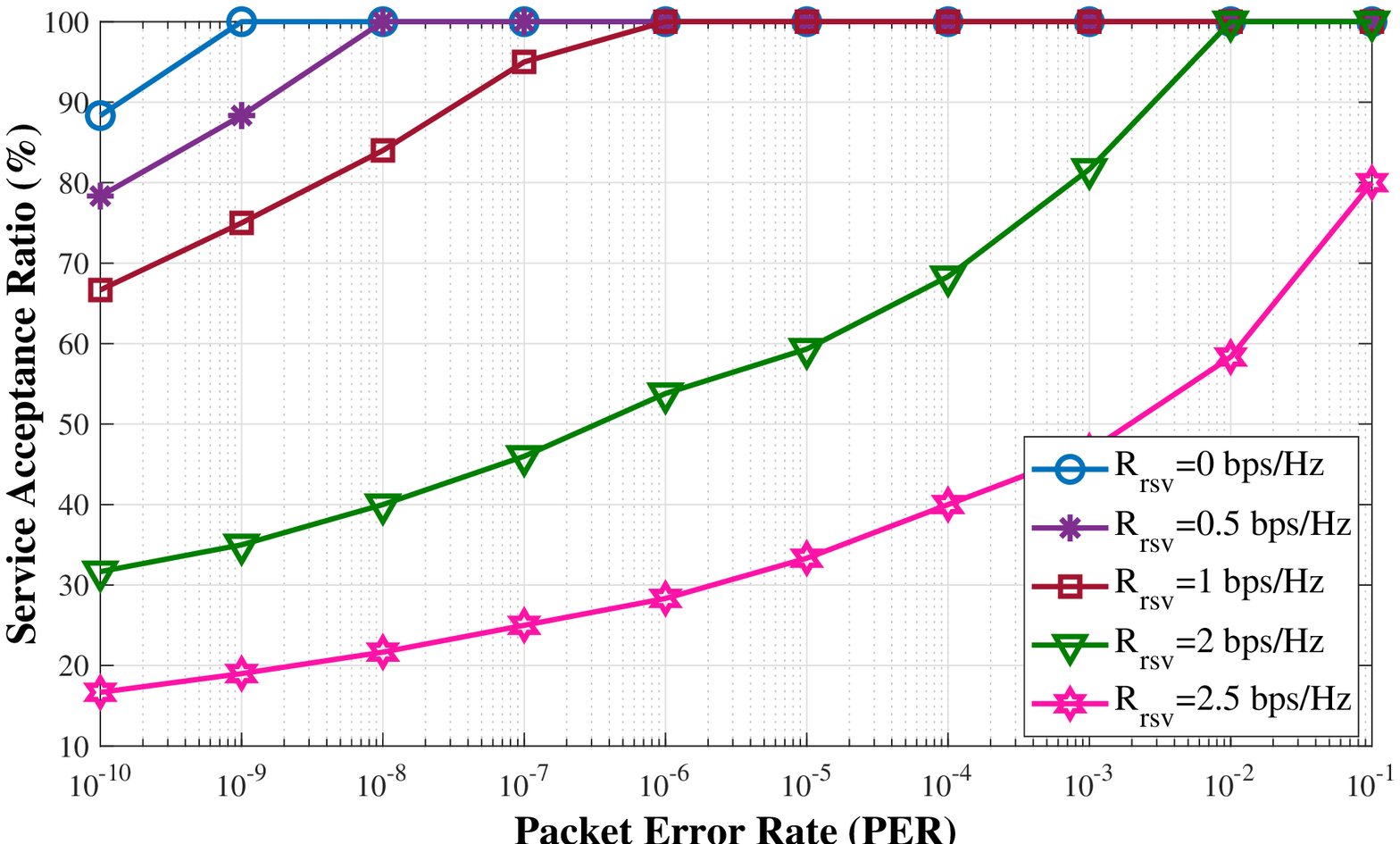}\label{Sim8}}	
	\vspace{-0.5em}
	\caption{System performance versus Packet error rate (PER)}
	\label{Sim78}
\end{figure*}

\begin{figure*}[t]
		\vspace{-1em}
	\centering
	\subfloat[Total transmit power vs. E2E delay]{\includegraphics[width=0.45\textwidth]{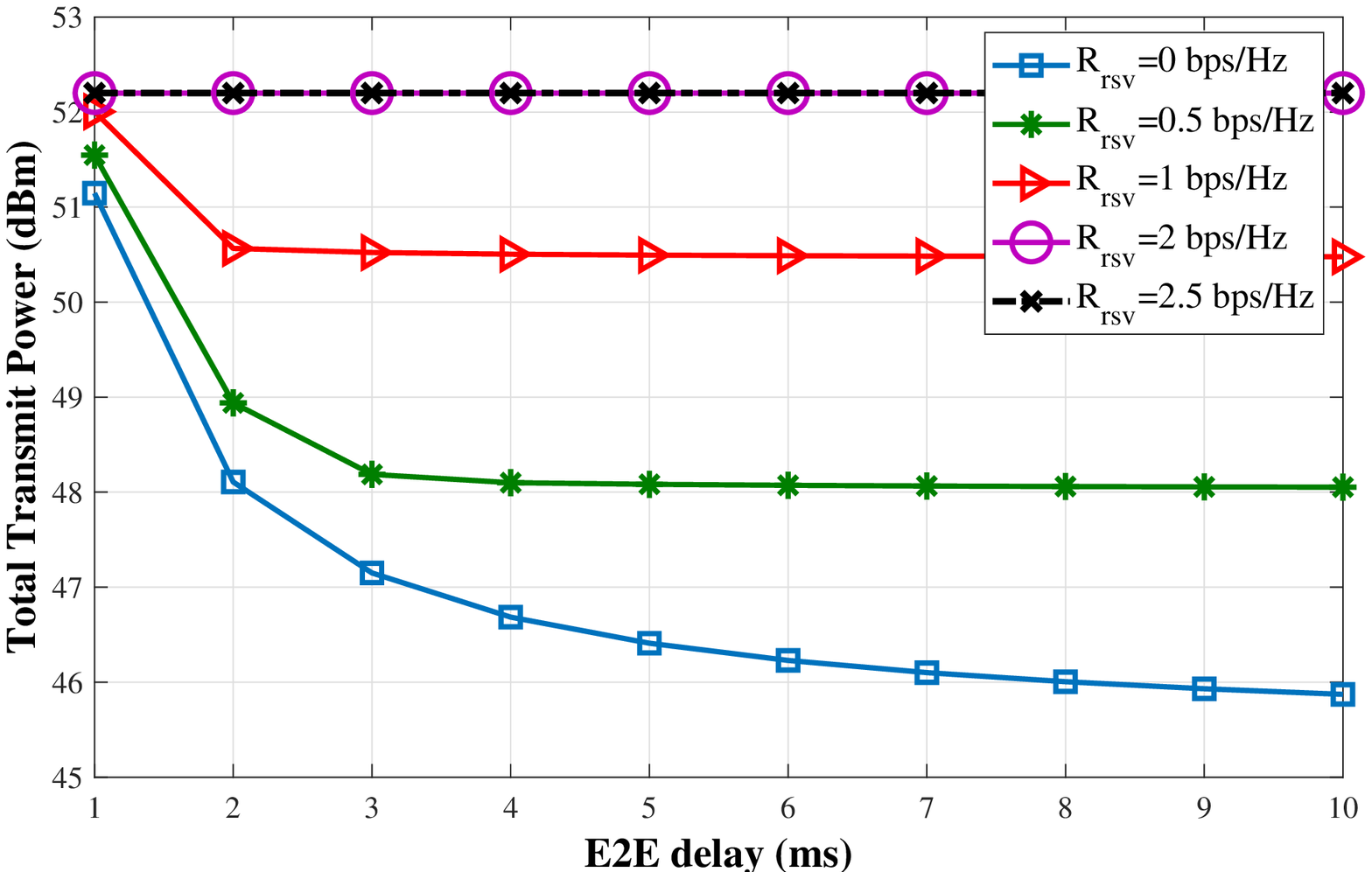}\label{Sim1}}
	\subfloat[Service acceptance ratio vs. E2E delay]{\includegraphics[width=0.45\textwidth]{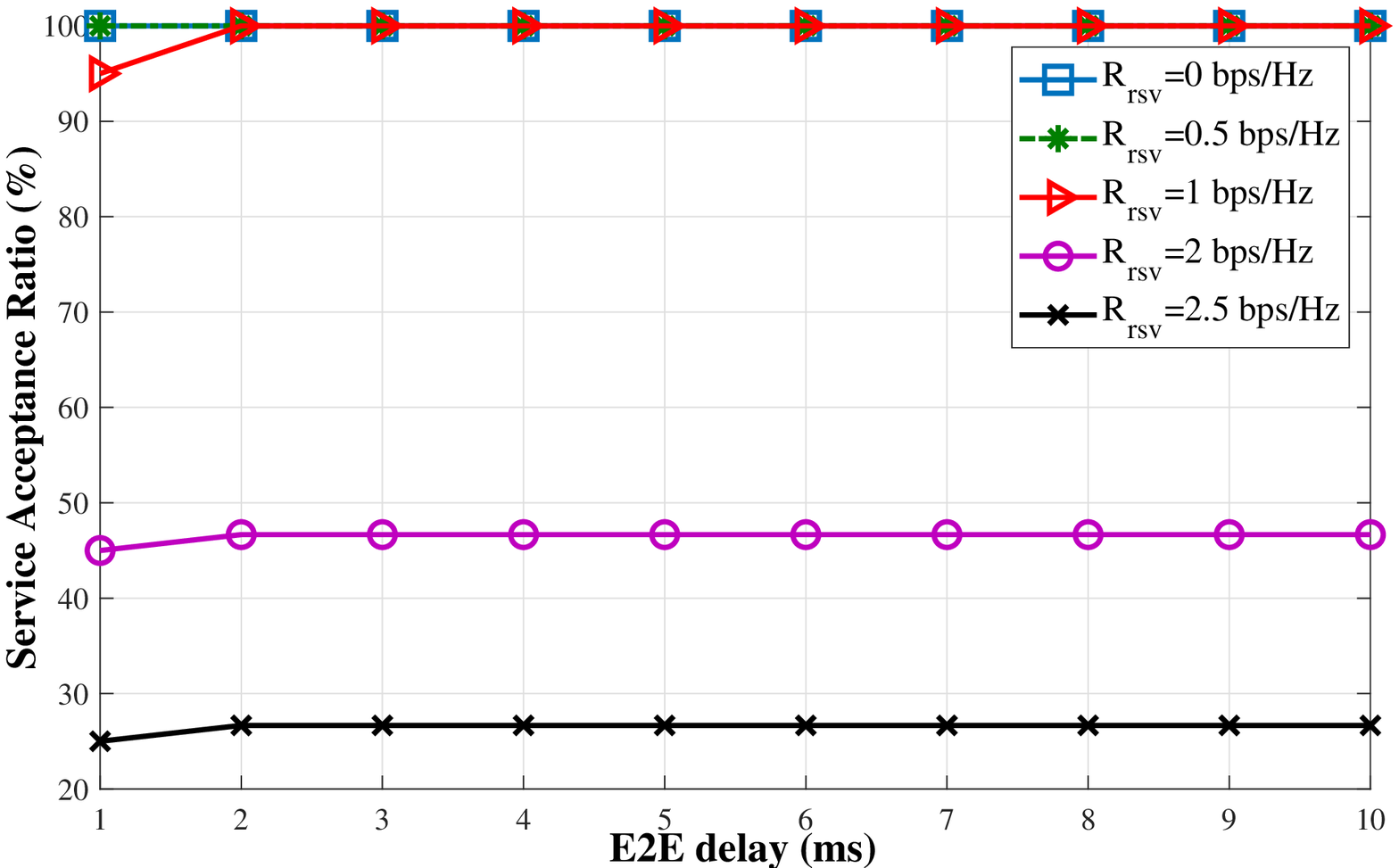}\label{Sim2}}	
	\vspace{-0.5em}
	\caption{System performance versus E2E delay}
	\label{Sim12}
\end{figure*}

Here, we investigate the impact of the delay and reliability on the system performance.
As can be seen from Figures \ref{Sim78} and \ref{Sim12},  for low PER (values close to $10^{-10}$), i.e., high reliability requirement and low E2E delay requirements (values close to $1$~ms), the total transmit power is high and the {SAR} is low. 
	 By relaxing the reliability requirement (PER close to $10^{-1}$) or delay requirement (E2E delay close to $10$~ms), the total transmit power decreases and also the {SAR} increases.
	  Therefore, for high reliability (low PER) or low E2E delay services, it is necessary to increase the amount of the network resources such as transmit power to reduce the number of rejected users.
	  Moreover, it can be seen that compared to the E2E delay, the reliability has more impact on the total transmit power and the {SAR}.
\vspace{-2em}
\subsection{Admission Control Performance}
\begin{figure}[t]
	\centering
	\includegraphics[scale=0.45]{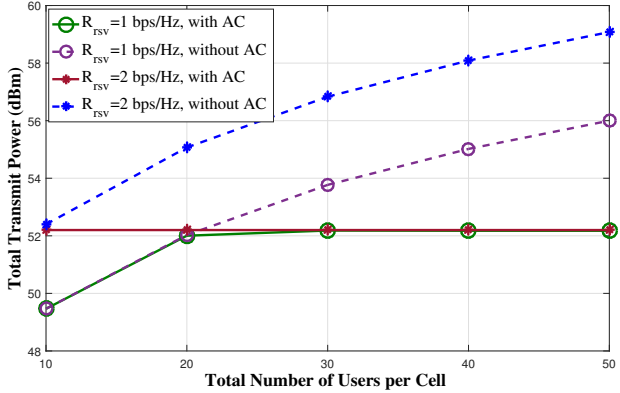}
	\vspace{-0.5em}
	\caption{Admission Control Performance}
	\label{SimAC}
\end{figure}

Here, we investigate the AC effect on the system performance and compare it with the case without AC. In this regard, we remove the power budget constraints from problem \eqref{eqoa} and solve the following problem:
	\begin{align}\label{eqoaWAC}
	&\mathop {\min }\limits_{_{\scriptstyle{\bf{P}},{\bf{T}},\hfill\atop
			\scriptstyle{\bf{X}},{\bf{D}}\hfill}} \sum\limits_{j \in {\cal J}} {\sum\limits_{{k_2} \in {{\cal K}_2}} {\sum\limits_{s \in {\cal S}} {\sum\limits_{u \in {{\cal I}_s}} {\sum\limits_{{k_1} \in {{\cal K}_1}} {\sum\limits_{q \in {\cal Q}} {x_{{k_2}}^{j,q}} } } } } } p_{{k_2}}^{j,q} + \tau _{u,{k_1}}^{s,j,q}p_{u,{k_1}}^{s,j,q}\nonumber
	\\\text{s.t.}:&\text{(C1)-(C3), (C6), (C9)-(C14)}, \nonumber
	\end{align}
	This new problem can be solved with the iterative algorithm and DC approximation. In problem  \eqref{eqoa} the power budget constraints, i.e., (C4)-(C5) and (C7)-(C8), restrict the amount of total transmit power, and if the power is not enough, the problem becomes infeasible. Therefore, with the AC method, some users are rejected and the problem becomes feasible. By removing power constraints, the total transmit power increases to satisfy all constraints. Thus, as shown in Fig. \ref{SimAC}, for $R_{\text{rsv}}=2$, the amount of the total transmit power without employing AC is increased by about $7$~dB.
\vspace{-1em}
\subsection{Dynamic Approach versus Fixed approach}
 To evaluate the performance of the proposed system model, we consider a different scenario which is called fixed approach. In our setup, we adjust the delay dynamically to minimize the transmit power which is called dynamic approach. In the traditional scenario, i.e., fixed approach, we assume that the delay constraints are fixed and cannot be adjusted for the access and fronthaul links. In this case, we have a new optimization problem, referred to as, the relaxed problem, in which we remove the delay variables from problem \eqref{In_val} and ignore constraint C9. Furthermore, we set access and fronthaul delays manually in constraints \~C10, \~C11, and \~C12 as ${D^j_{{\rm{max}}}}={D^{\text{BBU}}_{{\rm{max}}}}={D^{i,j}_{{\rm{max}}}}=D^{\text{max}}_{i,j,s}/3$. The new problem can be solved by DC approximation.
\begin{figure*}[t]
		\vspace{-2em}
	\centering
	\subfloat[Total transmit power vs. total number of users per cell]{\includegraphics[width=0.45\textwidth]{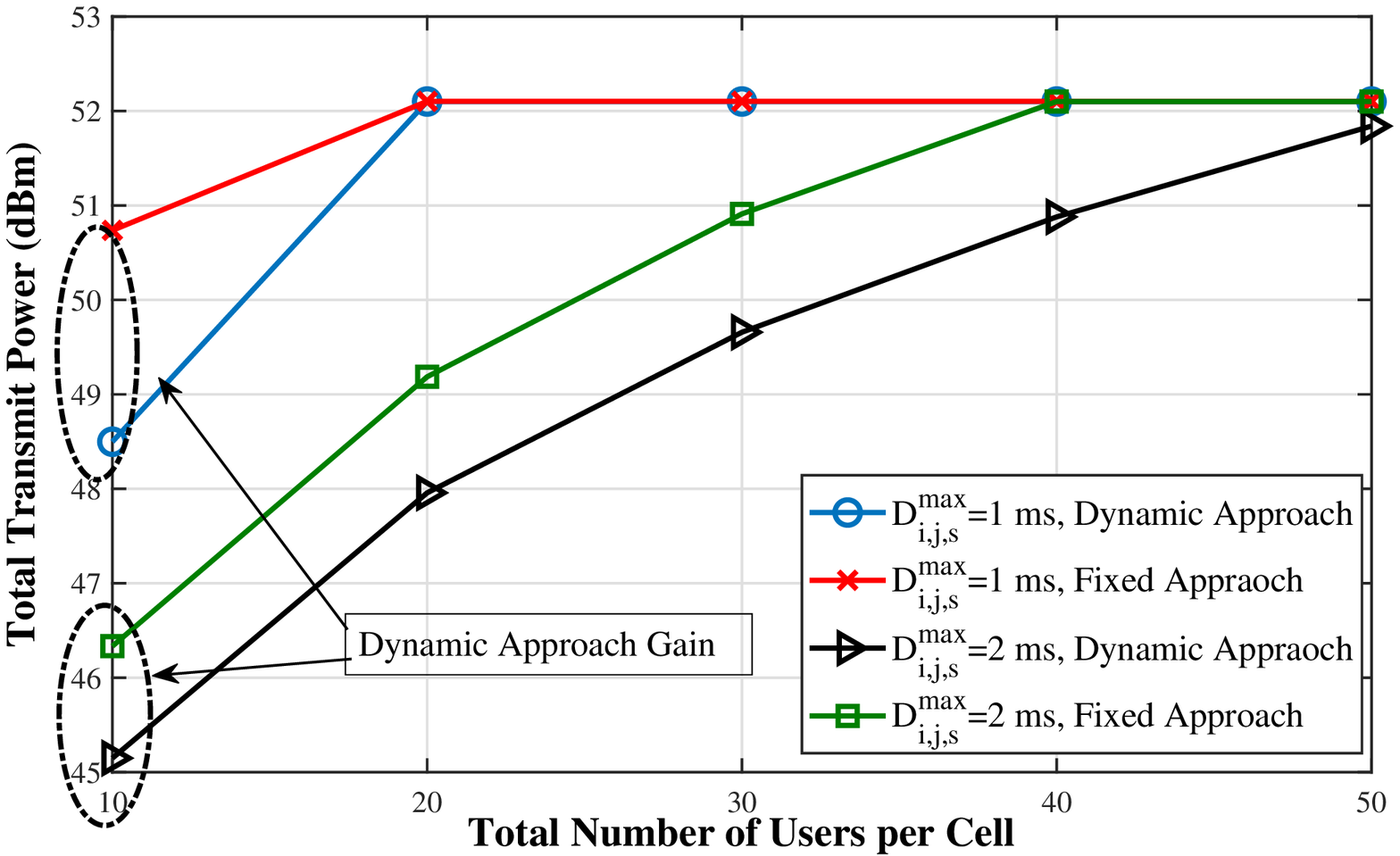}\label{Sim9}}
	\subfloat[Service acceptance ratio vs. total number of users per cell]{\includegraphics[width=0.45\textwidth]{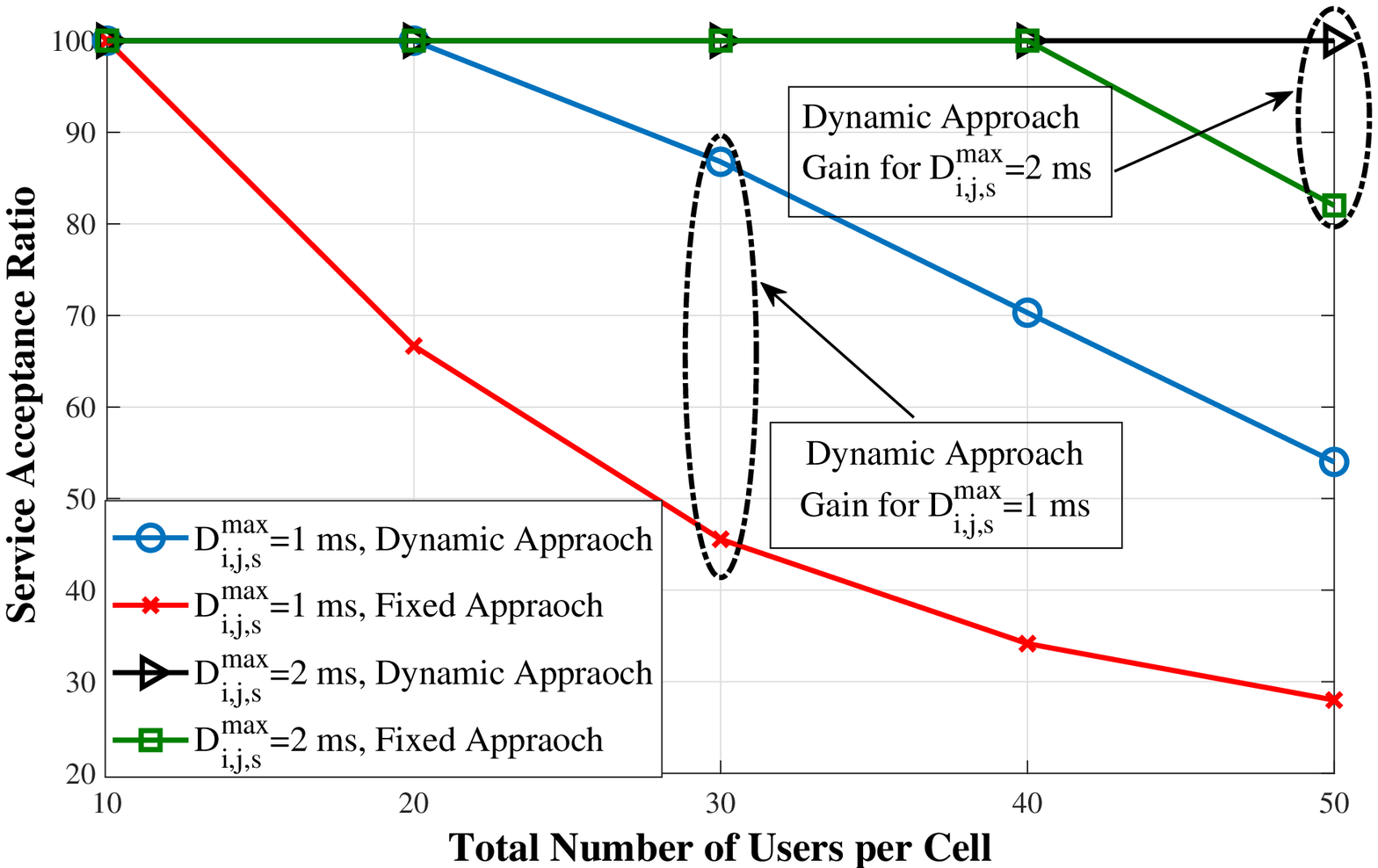}\label{Sim10}}	
	\vspace{-0.5em}
	\caption{Performance comparison of our proposed dynamic approach versus fixed approach}
	\label{Sim910}
\end{figure*}
In Fig. \ref{Sim9}, we investigate the effect of the actual value of the delay $D^{\text{max}}_{i,j,s}=1,2$~ms on the total transmit power.
It is evident that  for $1$~ms E2E delay and $10$ users, in the fixed approach compared to the dynamic approach, the total transmit power increases by $2$~dB.
	However, for a large number of users, the total transmit power is not enough and the acceptance ratio significantly decreases significantly in the fixed approach.  
 Moreover, from Fig. \ref{Sim910}, the proposed system model has a considerably better performance than the system model corresponding to the relaxed problem (fixed approach).
  As can be seen from \ref{Sim910}(a),  when we have enough power, by the dynamic adjustment of the delay, we can save around $2$~dB in transmit power. Furthermore, when the power is not enough e.g., for 30 users, we can see from Fig. \ref{Sim910}(b), in the fixed approach the {SAR} is below 50\% while 85\% of users are accepted in our proposed algorithm (dynamic approach).
\vspace{-1em}
\subsection{Convergence Study of Algorithm \ref{ALG1}}
\begin{figure}[t]
	\centering
	\includegraphics[scale=0.45]{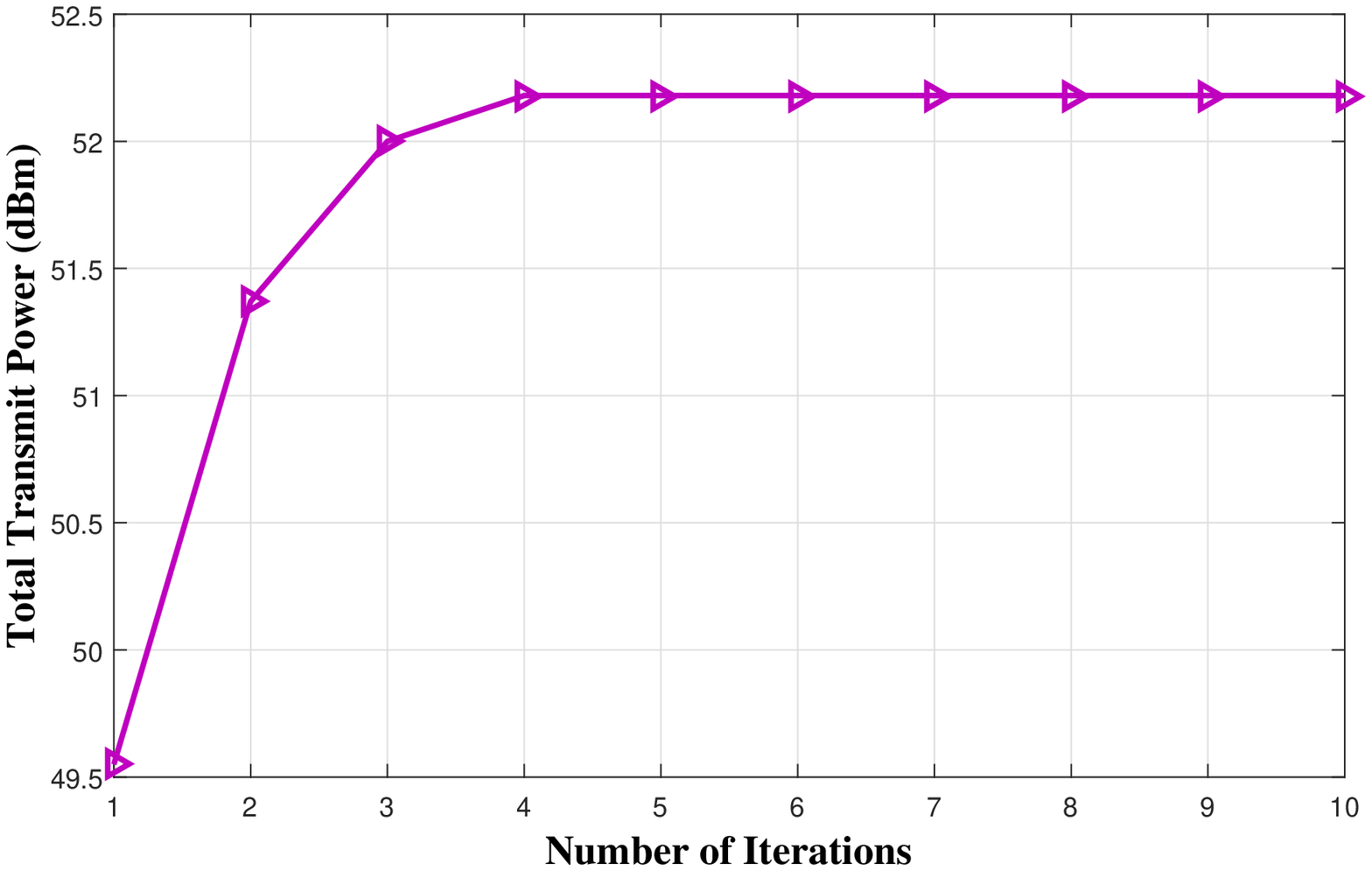}
	\vspace{-0.5em}
	\caption{Convergence of Algorithm.\ref{ALG1}}
	\label{Sim11}
\end{figure}

In this subsection, we investigate the proof of the proposed system model convergence in Propositions 1-3. In Fig. \ref{Sim11}, the convergence of the alternate method for the proposed system model is demonstrated.  It can be seen that the  solution of the proposed algorithm converges  to a fixed value after 15 iterations. For this simulation, we set $R_{\text{rsv}}=0$~ bps/Hz and the total number of users per cell is equal to $50$. 
\vspace{-1em}
\section{Conclusion}
In this paper, we proposed a novel queuing model based on admission control (AC) for the {TI} services in  OFDMA-based C-RANs serving several pairs of tactile users. For each pair of tactile users within C-RAN coverage area, our setup includes RRH and BBU queuing delays in one E2E connection which is a more practical scenario in this context compared to previous works. We proposed a resource allocation (RA) problem to minimize the transmit power by considering E2E delay and reliability of joint access and fronthaul links for each pair of tactile users  where the delays of fronthaul and access links are dynamically adjusted. 
We also propose how AC can be applied to convert infeasible situations of the system into feasible ones. 
	To solve the highly non-convex proposed RA problem, we applied the SCA method. Simulation results revealed that by dynamic adjustment of the access and fronthaul delays and admission control process, transmit power can be considerably saved and the service acceptance ratio can be significantly increased compared to the case of fixed approach.
\vspace{-1em}
\appendices
\section{}
	In this algorithm, the convergence can be guaranteed if we can show that the objective function is a decreasing function with respect to the number of iterations. For the algorithm in Table.\ref{ALG1}, in the first step of iteration $i+1$, with a given power allocation at iteration $i$, $\boldsymbol{x}=x^{(i+1)}$ and $\boldsymbol{\tau}=\tau^{(i+1)}$ are derived. Based on DC approximation, we will have $f(\boldsymbol{x^{(i)}},\boldsymbol{p^{(i)}})\le f(\boldsymbol{x^{(i+1)}},\boldsymbol{p^{(i)}})$ and  $f(\boldsymbol{\tau^{(i)}},\boldsymbol{p^{(i)}})\le f(\boldsymbol{\tau^{(i+1)}},\boldsymbol{p^{(i)}})$  \cite{ngo2014joint}. In the second step, with a given subcarrier allocation at iteration $i+1$, the power allocation at iteration $(i+1)$ is obtained. Based on DC approximation, we will have $f(\boldsymbol{x^{(i+1)}},\boldsymbol{p^{(i)}})\le f(\boldsymbol{x^{(i+1)}},\boldsymbol{p^{(i+1)}})$ and  $f(\boldsymbol{\tau^{(i+1)}},\boldsymbol{p^{(i)}})\le f(\boldsymbol{\tau^{(i+1)}},\boldsymbol{p^{(i+1)}})$. Finally, we have
$
	... \le f(\boldsymbol{x^{(i)}},\boldsymbol{p^{(i)}})\le  f(\boldsymbol{x^{(i+1)}},\boldsymbol{p^{(i)}}) \le f(\boldsymbol{x^{(i+1)}},\boldsymbol{p^{(i+1)}})\le ... \le f(\boldsymbol{x^{*}},\boldsymbol{p^{*}}),
	$
	and
	$
	... \le f(\boldsymbol{\tau^{(i)}},\boldsymbol{p^{(i)}})\le f(\boldsymbol{\tau^{(i+1)}},\boldsymbol{p^{(i)}}) \le f(\boldsymbol{\tau^{(i+1)}},\boldsymbol{p^{(i+1)}})\le ... \le f(\boldsymbol{\tau^{*}},\boldsymbol{p^{*}}),
	$
	where $x^{*}$,$\tau^{*}$ and $p^{*}$ are optimal solutions which are obtained in the previous iteration. After each iteration, we have $ f(\boldsymbol{\tau^{(i+1)}},\boldsymbol{p^{(i+1)}})-f(\boldsymbol{\tau^{(i)}},\boldsymbol{p^{(i)}})$ and  $ f(\boldsymbol{x^{(i+1)}},\boldsymbol{p^{(i+1)}})-f(\boldsymbol{x^{(i)}},\boldsymbol{p^{(i)}})$ which is a decreasing function and consequently the proposed algorithm converges.
	\vspace{-1em}
\section{}
	We approximate the rate function with the DC approximation as explained in Appendix C.
 In each iteration for each subproblem, 
 the objective function and all constraints are single variable functions. With 
  DC method which is described in Appendix C, the non-convex problem can be converted into a convex problem \cite{6108303}. Given the fact that the functions in each iteration for each subproblem are single-variable, we show the functions $f_{\text{AC}}^{q}(\boldsymbol{P},\boldsymbol{\tau}), g_{\text{AC}}^{q}(\boldsymbol{P},\boldsymbol{\tau}), y_{\text{AC}}^{q}(\boldsymbol{P},\boldsymbol{\tau}), 
  f_{\text{FH}}^{q}(\boldsymbol{P},\boldsymbol{x}),$ and $g_{\text{FH}}^{q}(\boldsymbol{P},\boldsymbol{x})$ as a function $\nu(\boldsymbol{\rho}),$ where according to the subproblem $\rho$ can be $\boldsymbol{P}$, $\boldsymbol{x}$ or $\boldsymbol{\tau}$  for simplicity. Therefore, we have
$
	\nu(\boldsymbol{\rho^{(i)}})\le \nu(\boldsymbol{\rho^{(i-1)}})+\nabla \nu(\boldsymbol{\rho^{(i-1)}})(\boldsymbol{\rho^{(i)}}-\boldsymbol{\rho^{(i-1)}}).
$	
	Consequently, from \cite{ngo2014joint}, for iteration $i$, we have
$
	f(\boldsymbol{\rho^{(i)}})-\{ g(\boldsymbol{\rho^{(i-1)}})+\nabla g(\boldsymbol{\rho^{(i-1)}})(\boldsymbol{\rho^{(i)}}-\boldsymbol{\rho^{(i-1)}})\}\ge R_0.
$	
	Moreover, we have
$
	f(\boldsymbol{\rho^{(i+1)}})- g(\boldsymbol{\rho^{(i+1)}})\ge f(\boldsymbol{\rho^{(i)}})- g(\boldsymbol{\rho^{(i)}}) -\nabla g(\boldsymbol{\rho^{(i)}})(\boldsymbol{\rho^{(i+1)}}-\boldsymbol{\rho^{(i)}})\ge f(\boldsymbol{\rho^{(i)}})- g(\boldsymbol{\rho^{(i)}})
$  \cite{6108303}. 
		In other words, after each iteration, a distance of new solution to the optimum solution is always smaller than that of the previous iteration \cite{6108303}.
		Therefore, SCA with the DC approximation converges to a suboptimal solution \cite{mokdad2016radio,6108303}.
\section{}
First, we transform the access rate into a convex function by using the DC approximation as equation  \eqref{EQAPXC},
\begin{figure*}
	\vspace{-1em}
	\begin{align}
	&\begin{array}{l}
\sum\limits_{s \in S} {\sum\limits_{u \in {I_s}} {\sum\limits_{j \in J} {\sum\limits_{{k_1} \in {{\cal K}_1}} {\tau _{u,{k_1}}^{s,j,{\rm{q}}}} r_{u,{k_1}}^{s,j,{\rm{q}}}} } }  = \sum\limits_{j \in J} {\sum\limits_{s \in S} {\sum\limits_{u \in {I_s}} {\sum\limits_{{k_1} \in {{\cal K}_1}} {\frac{{\tau _{u,{k_1}}^{s,j,{\rm{q}}}{w_{{k_1}}}}}{{\ln 2}}\bigg[\ln (1 + \gamma _{u,{k_1}}^{s,j,q}) - \sqrt {\frac{{V_{u,{k_1}}^{s,j,q}}}{{\phi {w_{{k_1}}}}}} f_Q^{ - 1}(\varepsilon _{u,{k_1}}^{s,j,q})\bigg]} } } } \\
\sum\limits_{j \in J} {\sum\limits_{s \in S} {\sum\limits_{u \in {I_s}} {\sum\limits_{{k_1} \in {{\cal K}_1}} {\frac{{\tau _{u,{k_1}}^{s,j,{\rm{q}}}{w_{{k_1}}}}}{{\ln 2}}\bigg[{{\ln }_2}(\frac{{\sigma _{u,{k_1}}^{s,j,q} + I_{u,{k_1}}^{s,j,q} + \tau _{u,{k_1}}^{s,j,q}p_{u,{k_1}}^{s,j,q}h_{i,{k_1}}^{s,j,q}}}{{\sigma _{u,{k_1}}^{s,j,q} + I_{u,{k_1}}^{s,j,q}}})\;\;} } } }  -
 \frac{1}{{\sqrt {\phi {w_{{k_1}}}} }}\sqrt {1 - \frac{1}{{{{(1 + \gamma _{i,{k_1}}^{s,j,q})}^2}}}} f_Q^{ - 1}(\varepsilon _{u,{k_1}}^{s,j,q})\bigg]=
\\ \sum\limits_{j \in J} {\sum\limits_{s \in S} {\sum\limits_{u \in {I_s}} {\sum\limits_{{k_1} \in {{\cal K}_1}} \bigg({\frac{{\tau _{u,{k_1}}^{s,j,{\rm{q}}}{w_{{k_1}}}}}{{\ln 2}}{{\ln }_2}(\sigma _{u,{k_1}}^{s,j,q} + I_{u,{k_1}}^{s,j,q} + \tau _{u,{k_1}}^{s,j,q}p_{u,{k_1}}^{s,j,q}h_{i,{k_1}}^{s,j,q})\;\;} } } }  - \frac{{\tau _{u,{k_1}}^{s,j,{\rm{q}}}{w_{{k_1}}}}}{{\ln 2}}{\ln _2}(\sigma _{u,{k_1}}^{s,j,q} + I_{u,{k_1}}^{s,j,q})
 -\\ \frac{{\tau _{u,{k_1}}^{s,j,{\rm{q}}}\sqrt {{w_{{k_1}}}} }}{{\ln 2\sqrt \phi  }}\sqrt {1 - \frac{1}{{{{(1 + \gamma _{i,{k_1}}^{s,j,q})}^2}}}} f_Q^{ - 1}(\varepsilon _{u,{k_1}}^{s,j,q})\bigg)=
\sum\limits_{j \in J} {\sum\limits_{s \in S} {\sum\limits_{u \in {I_s}} {\sum\limits_{{k_1} \in {{\cal K}_1}}  } } }f_{\text{AC}}^{q}(P_{u,{k_1}}^{s,j,q},\tau_{u,{k_1}}^{s,j,q}) - g_{\text{AC}}^{q}(P_{u,{k_1}}^{s,j,q},\tau_{u,{k_1}}^{s,j,q})-y_{\text{AC}}^{q}(P_{u,{k_1}}^{s,j,q},\tau_{u,{k_1}}^{s,j,q}),
\end{array}
	\end{align}\label{EQAPXC}
	\vspace{-0.5em}
	\hrule
	\end{figure*}
where $f_{\text{AC}}^{q}(P_{u,{k_1}}^{s,j,q},\tau_{u,{k_1}}^{s,j,q})$ and $g_{\text{AC}}^{q}(P_{u,{k_1}}^{s,j,q},\tau_{u,{k_1}}^{s,j,q})$ are concave functions as follows
\begin{equation*}
\begin{array}{l}
f_{\text{AC}}^{q}(P_{u,{k_1}}^{s,j,q},\tau_{u,{k_1}}^{s,j,q})=\frac{{{w_{{k_1}}}}}{{\ln 2}}{\ln _2}(\sigma _{u,{k_1}}^{s,j,q} + I_{u,{k_1}}^{s,j,q} + \tau _{u,{k_1}}^{s,j,q}p_{u,{k_1}}^{s,j,q}h_{i,{k_1}}^{s,j,q}),
\end{array}
\end{equation*}
\begin{equation*}
\begin{array}{l}
g_{\text{AC}}^{q}(P_{u,{k_1}}^{s,j,q},\tau_{u,{k_1}}^{s,j,q}) =\frac{{{w_{{k_1}}}}}{{\ln 2}} {\ln _2}(\sigma _{u,{k_1}}^{s,j,q} + I_{u,{k_1}}^{s,j,q}). \text{~~~~~~~~~~~~~~~~~~~~~}
\end{array}
\end{equation*}
  Then, to transform $\sum\limits_{s \in S} {\sum\limits_{u \in {I_s}} {\sum\limits_{j \in J} {\sum\limits_{{k_1} \in {{\cal K}_1}} {\tau _{u,{k_1}}^{s,j,{\rm{q}}}} r_{u,{k_1}}^{s,j,{\rm{q}}}} } }$ to a convex function, we use 
\begin{align}
\begin{array}{l}
{g_{\text{AC}}^{q}}(\boldsymbol{P},\boldsymbol{\tau}) \approx g_{{\rm{AC}}}^q{(\boldsymbol{P},\boldsymbol{\tau})^{({t_I} - 1)}} +\\ \nabla g{_{{\rm{AC}}}^q}^T  { (\boldsymbol{P},\boldsymbol{\tau}) ^{({t_I} - 1)}} ({(\boldsymbol{P},\boldsymbol{\tau})^{({t_I})}} - {(\boldsymbol{P},\boldsymbol{\tau})^{({t_I} - 1)}}),\text{~~~~~~~~~~~}
\end{array} \nonumber
\end{align}
 where $\nabla {g_{\text{AC}}^{q}}(\boldsymbol{P},\boldsymbol{\tau})$ for subcarrier allocation subproblem is as follows
\begin{equation}\nonumber
\begin{split} 
&\nabla {g_{\text{AC}}^{q}}(\boldsymbol{\tau}) = \left\{ {\begin{array}{*{20}{c}}
	0,&{{\rm{if~}}m = j},\\
	{\frac{{h_{u,{k_1}}^{s,m,q}}p_{u,{k_1}}^{s,j,q}}{{\sigma _{u,{k_1}}^{s,j,q} +  I_{u,{k_1}}^{s,j,q}}}},&{{\rm{if~}}m \ne j},
	\end{array}} \right.
\end{split}
\end{equation}
and for power allocation subproblem, we have
\begin{equation}\nonumber
\begin{split} 
&\nabla {g_{\text{AC}}^{q}}(\boldsymbol{P}) = \left\{ {\begin{array}{*{20}{c}}
	0,&{{\rm{if~}}m = j},\\
	{\frac{{h_{u,{k_1}}^{s,m,q}}\tau_{u,{k_1}}^{s,j,q}}{{\sigma _{u,{k_1}}^{s,j,q} +  I_{u,{k_1}}^{s,j,q}}}},&{{\rm{if~}}m \ne j}.
	\end{array}} \right.
\end{split}
\end{equation}

Similarity, for $ {y_{\text{AC}}^{q}}(P_{u,{k_1}}^{s,j,q},\tau_{u,{k_1}}^{s,j,q})=\frac{{\tau _{u,{k_1}}^{s,j,{\rm{q}}}\sqrt {{w_{{k_1}}}} }}{{\ln 2\sqrt \phi  }}$ $\sqrt {1 - \frac{1}{{{{(1 + \gamma _{i,{k_1}}^{s,j,q})}^2}}}} f_Q^{ - 1}(\varepsilon _{u,{k_1}}^{s,j,q})$, we have
\begin{align}
\begin{array}{l}
{y_{\text{AC}}^{q}}(\boldsymbol{P},\boldsymbol{\tau}) \approx y_{{\rm{AC}}}^q{(\boldsymbol{P},\boldsymbol{\tau})^{({t_I} - 1)}} +\\ \nabla y{_{{\rm{AC}}}^q}^T  { (\boldsymbol{P},\boldsymbol{\tau}) ^{({t_I} - 1)}} ({(\boldsymbol{P},\boldsymbol{\tau})^{({t_I})}} - {(\boldsymbol{P},\boldsymbol{\tau})^{({t_I} - 1)}}). \text{~~~~~~~~~~} 
\end{array}\nonumber
\end{align}

Notice that, in power allocation subproblem, the values of subcarrier allocation subproblem are fixed, and in subcarrier allocation subproblem, the variables of transmit power allocations are fixed.
 Therefore, in power allocation subproblem, $\nabla {y_{\text{AC}}^{q}}(\boldsymbol{P})$ is calculated in equation \eqref{eqwer}
\begin{figure*}
	\vspace{-1em}
	\begin{equation} \label{eqwer}
	\begin{split} 
	&\nabla {y_{\text{AC}}^{q}}(\boldsymbol{P}) = \left\{ {\begin{array}{*{20}{c}}
		{\frac{{0.5(2\tau _{u,{k_1}}^{s,j,q}p_{u,{k_1}}^{s,j,q}|h_{u,{k_1}}^{s,j,q}{|^2} + 2\sigma _{u,{k_1}}^{s,j,q}\tau _{u,{k_1}}^{s,j,q}h_{u,{k_1}}^{s,j,q} + 2I_{u,{k_1}}^{s,j,q}\tau _{u,{k_1}}^{s,j,q}h_{u,{k_1}}^{s,j,q}){{(\psi _{u,{k_1}}^{s,j,q})}^{ - 1}}(\Gamma _{u,{k_1}}^{s,j,q}) - (\tau _{u,{k_1}}^{s,j,q}h_{u,{k_1}}^{s,j,q})\psi _{u,{k_1}}^{s,j,q}}}{{{{(\Gamma _{u,{k_1}}^{s,j,q} )}^2}}}},&{{\rm{if~}}m = j},\\
		\frac{{0.5(2\tau _{u,{k_1}}^{s,m,q}h_{u,{k_1}}^{s,m,q}p_{u,{k_1}}^{s,j,q}\tau _{u,{k_1}}^{s,j,q}h_{u,{k_1}}^{s,j,q}){{(\psi _{u,{k_1}}^{s,j,q})}^{ - 1}}(\Gamma _{u,{k_1}}^{s,j,q}) - (\tau _{u,{k_1}}^{s,m,q}h_{u,{k_1}}^{s,m,q})\psi _{u,{k_1}}^{s,j,q}}}{{{{(\Gamma _{u,{k_1}}^{s,j,q} )}^2}}}~~~~~~~~~~~~~~~~~~~~~,&{{\rm{if~}}m \ne j}.
		\end{array}} \right.
	\end{split}
	\end{equation}
	\vspace{-0.5em}
		\hrule
\end{figure*}
in which \\ $\psi _{u,{k_1}}^{s,j,q} =$$ \sqrt {p_{u,{k_1}}^{s,j,q}\tau _{u,{k_1}}^{s,j,q}h_{u,{k_1}}^{s,j,q}(p_{u,{k_1}}^{s,j,q}\tau _{u,{k_1}}^{s,j,q}h_{u,{k_1}}^{s,j,q} + 2\sigma _{u,{k_1}}^{s,j,q} + 2I_{u,{k_1}}^{s,j,q})}$ and $\Gamma _{u,{k_1}}^{s,j,q} = \sigma _{u,{k_1}}^{s,j,q} + I_{u,{k_1}}^{s,j,q} + \tau _{u,{k_1}}^{s,j,q}p_{u,{k_1}}^{s,j,q}h_{u,{k_1}}^{s,j,q}$.
Similarly, in subcarrier allocation subproblem, $\nabla {y_{\text{AC}}^{q}}(\boldsymbol{\tau})$ is calculated as \eqref{eqwasdf}.
\begin{figure*}
\begin{equation}\label{eqwasdf}
\begin{split} 
&\nabla {y_{\text{AC}}^{q}}(\boldsymbol{\tau}) = \left\{ {\begin{array}{*{20}{c}}
	{\frac{{0.5(2\tau _{u,{k_1}}^{s,j,q}|p_{u,{k_1}}^{s,j,q}h_{u,{k_1}}^{s,j,q}{|^2} + 2\sigma _{u,{k_1}}^{s,j,q}p_{u,{k_1}}^{s,j,q}h_{u,{k_1}}^{s,j,q} + 2I_{u,{k_1}}^{s,j,q}p_{u,{k_1}}^{s,j,q}h_{u,{k_1}}^{s,j,q}){{(\psi _{u,{k_1}}^{s,j,q})}^{ - 1}}(\Gamma _{u,{k_1}}^{s,j,q}) - (p_{u,{k_1}}^{s,j,q}h_{u,{k_1}}^{s,j,q})\psi _{u,{k_1}}^{s,j,q}}}{{{{(\Gamma _{u,{k_1}}^{s,j,q})}^2}}}},&{{\rm{if~}}m = j},\\
	\frac{{0.5(2 p _{u,{k_1}}^{s,m,q}h_{u,{k_1}}^{s,m,q}p_{u,{k_1}}^{s,j,q}\tau _{u,{k_1}}^{s,j,q}h_{u,{k_1}}^{s,j,q}){{(\psi _{u,{k_1}}^{s,j,q})}^{ - 1}}(\Gamma _{u,{k_1}}^{s,j,q}) - (p _{u,{k_1}}^{s,m,q}h_{u,{k_1}}^{s,m,q})\psi _{u,{k_1}}^{s,j,q}}}{{{{(\Gamma _{u,{k_1}}^{s,j,q} )}^2}}}~~~~~~~~~~~~~~~~~~~~~,&{{\rm{if~}}m \ne j}.
	\end{array}} \right.
\end{split}
\end{equation}
\vspace{-0.5em}
	\hrule
\end{figure*}
For fronthaul links, we have
\begin{align} \nonumber
	\begin{array}{l}
	\sum\limits_{j \in {\cal J}} {\sum\limits_{{k_2} \in {{\cal K}_2}} {x_{{k_2}}^{j,q}} } r_{{k_2}}^{j,q} = \sum\limits_{j \in J} {\sum\limits_{{k_2} \in {K_2}} {x_{{k_2}}^{j,q}} } \frac{{{w_{{k_2}}}}}{{\ln 2}}\left[ {\ln (1 + \frac{{p_{{k_2}}^{j,q}h_{{k_2}}^{j,q}}}{{\sigma _{{k_2}}^{j,q}}})} \right. - \\
	{\rm{ }}\left. {\sqrt {\frac{{V_{{k_2}}^{j,q}}}{{\phi {w_{{k_2}}}}}} f_Q^{ - 1}(\varepsilon _{{k_2}}^{j,q})} \right]+
	\sum\limits_{j \in {\cal J}} {\sum\limits_{{k_2} \in {{\cal K}_2}} {\frac{{x_{{k_2}}^{j,q}{w_{{k_2}}}}}{{\ln 2}}\ln (1 + \frac{{p_{{k_2}}^{j,q}h_{{k_2}}^{j,q}}}{{\sigma _{{k_2}}^{j,q}}}) }}-\\ \frac{{x_{{k_2}}^{j,q}\sqrt {{w_{{k_2}}}} }}{{\ln 2\sqrt \phi  }}\sqrt {1 - \frac{1}{{{{(1 + \gamma _{{k_2}}^{j,q})}^2}}}} f_Q^{ - 1}(\varepsilon _{{k_2}}^{j,q})  =\\\sum\limits_{j \in {\cal J}} {\sum\limits_{{k_2} \in {{\cal K}_2}}}f_{\text{FH}}^{q}(p _{{k_2}}^{j,q},x_{{k_2}}^{j,q}) - g_{\text{FH}}^{q}(p _{{k_2}}^{j,q},x_{{k_2}}^{j,q}),
	\end{array}
\end{align}
where $f_{\text{FH}}^{q}(p _{{k_2}}^{j,q},x_{{k_2}}^{j,q})=\frac{{x_{{k_2}}^{j,q}{w_{{k_2}}}}}{{\ln 2}}\ln (1 + \frac{{p_{{k_2}}^{j,q}h_{{k_2}}^{j,q}}}{{\sigma _{{k_2}}^{j,q}}}) $ and $g_{\text{FH}}^{q}(p _{{k_2}}^{j,q},x_{{k_2}}^{j,q})=\frac{{x_{{k_2}}^{j,q}\sqrt {{w_{{k_2}}}} }}{{\ln 2\sqrt \phi  }}\sqrt {1 - \frac{1}{{{{(1 + \gamma _{{k_2}}^{j,q})}^2}}}} f_Q^{ - 1}(\varepsilon _{{k_2}}^{j,q})$.
Then, we deploy $
{g_{\text{FH}}^{q}}(\boldsymbol{P},\boldsymbol{X}) \approx {g_{\text{FH}}^{q}}(\boldsymbol{P},\boldsymbol{X})^{({t_I} - 1)} + \nabla {g_{\text{FH}}^{q}}(\boldsymbol{P},\boldsymbol{X})^{({t_I} - 1)}((\boldsymbol{P},\boldsymbol{X})^{({t_I})} - (\boldsymbol{P},\boldsymbol{X})^{({t_I} - 1)}),
$ in which for power subcarrier subproblem, we have
\begin{equation}\nonumber
\begin{array}{l}
\nabla g_{{\rm{FH}}}^q({\bf{P}}) = \frac{{0.5(2x_{{k_2}}^{j,q}p_{{k_2}}^{j,q}|h_{{k_2}}^{j,q}{|^2} + 2\sigma _{{k_2}}^{j,q}x_{{k_2}}^{j,q}h_{{k_2}}^{j,q}){{(\Phi _{u,{k_1}}^{s,j,q})}^{ - 1}}(\Pi _{u,{k_1}}^{s,j,q})}}{{{{(\Pi _{u,{k_1}}^{s,j,q})}^2}}} - \\
\frac{{ x_{{k_2}}^{j,q}h_{{k_2}}^{j,q}(\Phi _{u,{k_1}}^{s,j,q})}}{{{{(\Pi _{u,{k_1}}^{s,j,q})}^2}}},
\end{array}
\end{equation}
where $\Phi _{u,{k_1}}^{s,j,q} = \sqrt {x_{{k_2}}^{j,q}p_{{k_2}}^{j,q}h_{{k_2}}^{j,q}(x_{{k_2}}^{j,q}p_{{k_2}}^{j,q}h_{{k_2}}^{j,q} + 2\sigma _{{k_2}}^{j,q}x_{{k_2}}^{j,q}p_{{k_2}}^{j,q}h_{{k_2}}^{j,q})}$ and $\Pi _{u,{k_1}}^{s,j,q} = \sigma _{{k_2}}^{j,q} + x_{{k_2}}^{j,q}p_{{k_2}}^{j,q}h_{{k_2}}^{j,q}$.
For subcarrier allocation subproblem, we have
\begin{equation}\nonumber
\begin{array}{l}
\nabla g_{{\rm{FH}}}^q({\bf{X}}) = \frac{{0.5(2x_{{k_2}}^{j,q}|p_{{k_2}}^{j,q}h_{{k_2}}^{j,q}{|^2} + 2\sigma _{{k_2}}^{j,q}p_{{k_2}}^{j,q}h_{{k_2}}^{j,q}){{(\Phi _{u,{k_1}}^{s,j,q})}^{ - 1}}(\Pi _{u,{k_1}}^{s,j,q})}}{{{{(\Pi _{u,{k_1}}^{s,j,q})}^2}}} - \\
\frac{{p_{{k_2}}^{j,q}h_{{k_2}}^{j,q}(\Phi _{u,{k_1}}^{s,j,q})}}{{{{(\Pi _{u,{k_1}}^{s,j,q})}^2}}}
\end{array}
\end{equation}
 C13 and C14 are functions of the rates, i.e., \eqref{eqo3} and \eqref{eqo10}. Therefore, by the DC approximation, C13 is transformed into a convex function in equation \eqref{C13C14Trans}, 
 \begin{figure*}
 	\vspace{-1em}
\begin{equation}\label{C13C14Trans}
\begin{array}{l}
-{\sum\limits_{{s} \in {{\cal S}}}  \sum\limits_{u \in {\cal I}_s} {\sum\limits_{j \in {\cal J}} {\sum\limits_{{k_1} \in {{\cal K}_1}}}}} {\tau_{u,k_1}^{s,j,\text{UL}} r_{u,{k_1}}^{{s,j},UL}}+{\sum\limits_{j \in {\cal J}} {\sum\limits_{{k_2} \in {{\cal K}_2}}}} {x_{k_2}^{j,\text{UL}}r_{{k_2}}^{{j},{\rm{UL}}}}  =
\\\sum\limits_{j \in J} {\sum\limits_{s \in S} {\sum\limits_{u \in {I_s}} {\sum\limits_{{k_1} \in {K_1}} \bigg (  - \frac{{\tau _{u,{k_1}}^{s,j,{\rm{q}}}{w_{{k_1}}}}}{{\ln 2}}{{\ln }_2}(\sigma _{u,{k_1}}^{s,j,q} + I_{u,{k_1}}^{s,j,q} + \tau _{u,{k_1}}^{s,j,q}p_{u,{k_1}}^{s,j,q}h_{i,{k_1}}^{s,j,q})\;\;} } }  +\\ \frac{{\tau _{u,{k_1}}^{s,j,{\rm{q}}}{w_{{k_1}}}}}{{\ln 2}}{\ln _2}(\sigma _{u,{k_1}}^{s,j,q} + I_{u,{k_1}}^{s,j,q}) +\frac{{\tau _{u,{k_1}}^{s,j,{\rm{q}}}\sqrt {{w_{{k_1}}}} }}{{\ln 2\sqrt \phi  }}\sqrt {1 - \frac{1}{{{{(1 + \gamma _{i,{k_1}}^{s,j,q})}^2}}}} f_Q^{ - 1}(\varepsilon _{u,{k_1}}^{s,j,q})\bigg)+\sum\limits_{j \in J} {\sum\limits_{{k_2} \in {K_2}}\bigg( {\frac{{x_{{k_2}}^{j,q}{w_{{k_2}}}}}{{\ln 2}}\ln (1 + \frac{{p_{{k_2}}^{j,q}h_{{k_2}}^{j,q}}}{{\sigma _{{k_2}}^{j,q}}})}} -\\ \frac{{x_{{k_2}}^{j,q}\sqrt {{w_{{k_2}}}} }}{{\ln 2\sqrt \phi  }}\sqrt {1 - \frac{1}{{{{(1 + \gamma _{{k_2}}^{j,q})}^2}}}} f_Q^{ - 1}(\varepsilon _{{k_2}}^{j,q})\bigg) ={\sum\limits_{{s} \in {{\cal S}}}  \sum\limits_{u \in {\cal I}_s} {\sum\limits_{j \in {\cal J}} {\sum\limits_{{k_1} \in {{\cal K}_1}}}}}\bigg (- f_{\text{AC}}^{\text{UL}}( p_{u,{k_1}}^{{s,j},q}, \tau_{u,{k_1}}^{{s,j},q})+ g_{\text{AC}}^{\text{UL}}( p_{u,{k_1}}^{{s,j},q}, \tau_{u,{k_1}}^{{s,j},q})\\+y_{{\rm{AC}}}^q(P_{u,{k_1}}^{s,j,q},\tau _{u,{k_1}}^{s,j,q})\bigg)+ {\sum\limits_{j \in {\cal J}} {\sum\limits_{{k_2} \in {{\cal K}_2}}}}f_{\text{FH}}^{q}(p _{{k_2}}^{j,q},x_{{k_2}}^{j,q}) - g_{\text{FH}}^{q}(p _{{k_2}}^{j,q},x_{{k_2}}^{j,q}),
\end{array}
\end{equation}
	\vspace{-0.5em}
	\hrule
 \end{figure*} 
where  $- f_{\text{AC}}^{\text{UL}}( p_{u,{k_1}}^{{s,j},q}, \tau_{u,{k_1}}^{{s,j},q})$ is non-convex and based on DC approximation, we convert it to a convex function as follows:
\begin{align}
\begin{array}{l}
f_{{\rm{AC}}}^{{\rm{UL}}}(\boldsymbol{P},\boldsymbol{\tau}) \approx  {\rm{ }}f_{{\rm{AC}}}^{{\rm{UL}}}{(\boldsymbol{P},\boldsymbol{\tau})^{({t_I} - 1)}} +\\ f_{{\rm{AC}}}^{{\rm{UL}}}{(\boldsymbol{P},\boldsymbol{\tau})^{({t_I} - 1)}}({(\boldsymbol{P},\boldsymbol{\tau})^{({t_I})}} - {(\boldsymbol{P},\boldsymbol{\tau})^{({t_I} - 1)}}),
\end{array}  \nonumber
\end{align}
in which for subcarrier allocation subproblem, we have
\begin{equation}
\begin{split}
&\nabla f_{\text{AC}}^{\text{UL}}(\boldsymbol{\tau})= \left\{ {\begin{array}{*{20}{c}}
	{\frac{{h_{u,{k_1}}^{j,m,\rm{UL}}}{p_{u,{k_1}}^{s,j,\rm{UL}}}}{{\sigma _{u,{k_1}}^{s,j,\rm{UL}} +  I_{u,{k_1}}^{s,j,\rm{UL}}}}}&{{\rm{if~~}}m =j},\\
	{\frac{{h_{u,{k_1}}^{s,m,\rm{UL}}}{p_{u,{k_1}}^{s,m,\rm{UL}}}}{{\sigma _{u,{k_1}}^{s,j,\rm{UL}} +  I_{u,{k_1}}^{s,j,\rm{UL}}}}}&{{\rm{if~~}}m \ne j},
	\end{array}} \right.
\end{split}
\end{equation}
and for power allocation subproblem, we have
\begin{equation}
\begin{split}
&\nabla f_{\text{AC}}^{\text{UL}}(\boldsymbol{P})= \left\{ {\begin{array}{*{20}{c}}
	{\frac{{h_{u,{k_1}}^{j,m,\rm{UL}}}{p_{u,{k_1}}^{s,j,\rm{UL}}}}{{\sigma _{u,{k_1}}^{s,j,\rm{UL}} +  I_{u,{k_1}}^{s,j,\rm{UL}}}}}&{{\rm{if~~}}m =j},\\
	{\frac{{h_{u,{k_1}}^{s,m,\rm{UL}}}{\tau_{u,{k_1}}^{s,m,\rm{UL}}}}{{\sigma _{u,{k_1}}^{s,j,\rm{UL}} +  I_{u,{k_1}}^{s,j,\rm{UL}}}}}&{{\rm{if~~}}m \ne j}.
	\end{array}} \right.
\end{split}
\end{equation}
Hence, C13 becomes a convex constraint via the DC approximation. Similarly, for C14, the rate functions are calculated in eqaution \eqref{C14END}
\begin{figure*}[t]
	\begin{equation}\label{C14END}
	\begin{array}{l}
	\sum \limits_{{s} \in {{ \mathcal{S}}}}  \sum \limits_{u \in {\mathcal{I}}_s} {\sum\limits_{j \in {\mathcal{J}}} {\sum \limits_{{k_1} \in {{\mathcal{K}}_1}}}} {\tau_{u,k_1}^{s,j,\text{UL}} r_{u,{k_1}}^{{s,j},UL}}-{\sum\limits_{j \in {\cal J}} {\sum\limits_{{k_2} \in {{\cal K}_2}}}} {x_{k_2}^{j,\text{UL}}r_{{k_2}}^{{j},{\rm{UL}}}}  =
	\sum\limits_{j \in J} {\sum\limits_{s \in S} {\sum\limits_{\scriptstyle u \in \hfill\atop
				\scriptstyle{I_s}\hfill} {\sum\limits_{\scriptstyle{k_1} \in \hfill\atop
					\scriptstyle{K_1}\hfill} \bigg( \frac{{\tau _{u,{k_1}}^{s,j,{\rm{q}}}{w_{{k_1}}}}}{{\ln 2}}{{\ln }_2}(\sigma _{u,{k_1}}^{s,j,q} + I_{u,{k_1}}^{s,j,q} + \tau _{u,{k_1}}^{s,j,q}p_{u,{k_1}}^{s,j,q}h_{i,{k_1}}^{s,j,q})\;} } }
	\\ 
	- \frac{{\tau _{u,{k_1}}^{s,j,{\rm{q}}}{w_{{k_1}}}}}{{\ln 2}}{\ln _2}(\sigma _{u,{k_1}}^{s,j,q} + I_{u,{k_1}}^{s,j,q}) - \frac{{\tau _{u,{k_1}}^{s,j,{\rm{q}}}\sqrt {{w_{{k_1}}}} }}{{\ln 2\sqrt \phi  }}f_Q^{ - 1}(\varepsilon _{u,{k_1}}^{s,j,q})\times \sqrt {1 - \frac{1}{{{{(1 + \gamma _{i,{k_1}}^{s,j,q})}^2}}}} \bigg)-\sum\limits_{j \in J} {\sum\limits_{{k_2} \in {K_2}}\bigg( {\frac{{x_{{k_2}}^{j,q}{w_{{k_2}}}}}{{\ln 2}}\ln (1 + \frac{{p_{{k_2}}^{j,q}h_{{k_2}}^{j,q}}}{{\sigma _{{k_2}}^{j,q}}})}} 
	\\
	+ \frac{{x_{{k_2}}^{j,q}\sqrt {{w_{{k_2}}}} }}{{\ln 2\sqrt \phi  }}\sqrt {1 - \frac{1}{{{{(1 + \gamma _{{k_2}}^{j,q})}^2}}}} f_Q^{ - 1}(\varepsilon _{{k_2}}^{j,q})\bigg) =
	{\sum\limits_{{s} \in {{\cal S}}}  \sum\limits_{u \in {\cal I}_s} {\sum\limits_{j \in {\cal J}} {\sum\limits_{{k_1} \in {{\cal K}_1}}}}}\bigg ( f_{\text{AC}}^{\text{UL}}( p_{u,{k_1}}^{{s,j},q}, \tau_{u,{k_1}}^{{s,j},q})- g_{\text{AC}}^{\text{UL}}( p_{u,{k_1}}^{{s,j},q}, \tau_{u,{k_1}}^{{s,j},q})

	-y_{{\rm{AC}}}^q(P_{u,{k_1}}^{s,j,q},\tau _{u,{k_1}}^{s,j,q})\bigg)\\+ {\sum\limits_{j \in {\cal J}} {\sum\limits_{{k_2} \in {{\cal K}_2}}}}-f_{\text{FH}}^{q}(p _{{k_2}}^{j,q},x_{{k_2}}^{j,q}) + g_{\text{FH}}^{q}(p _{{k_2}}^{j,q},x_{{k_2}}^{j,q}),
	\end{array}
	\end{equation}
		\hrule
\end{figure*}
where  $f_{\text{FH}}^{\text{DL}}({\bf{{{ P},{\bf{X}}}}})$ can be transformed to a convex function by DC approximation as follows
\begin{align}
\begin{array}{l}
 f_{{\rm{FH}}}^{{\rm{DL}}}(\boldsymbol{P},\boldsymbol{X}) \approx f_{{\rm{FH}}}^{{\rm{DL}}}{(\boldsymbol{P},\boldsymbol{X})^{({t_I} - 1)}} +\\ \nabla f_{{\rm{FH}}}^{{\rm{DL}}}{(\boldsymbol{P},\boldsymbol{X})^{({t_I} - 1)}}({(\boldsymbol{P},\boldsymbol{X})^{({t_I})}} - {(\boldsymbol{P},\boldsymbol{X})^{({t_I} - 1)}}),
\end{array}
\nonumber
\end{align}
 in which for subcarrier allocation subproblem, we have
 \begin{equation}
\begin{split} 
&\nabla f_{\text{FH}}^{\text{DL}}(\boldsymbol{X}) = \left\{ {\begin{array}{*{20}{c}}
	0,&{{\rm{if~}}m=j},\\
	{\frac{p_{{k_2}}^{j,\rm{DL}}{h_{{k_2}}^{m,\rm{DL}}}}{{\sigma _{{k_2}}^{j,\rm{DL}}  + p_{{k_2}}^{j,\rm{DL}}h_{{k_2}}^{j,\rm{DL}}x_{{k_2}}^{j,\rm{DL}}}}},&{{\rm{if~} m \neq j}},
	\end{array}} \right.
\end{split}
\end{equation}
and for power allocation subproblem, we have
 \begin{equation}
\begin{split} 
&\nabla f_{\text{FH}}^{\text{DL}}(\boldsymbol{P}) = \left\{ {\begin{array}{*{20}{c}}
	0,&{{\rm{if~} m=j}},\\
	{\frac{x_{{k_2}}^{j,\rm{DL}}{h_{{k_2}}^{m,\rm{DL}}}}{{\sigma _{{k_2}}^{j,\rm{DL}}  + p_{{k_2}}^{j,\rm{DL}}h_{{k_2}}^{j,\rm{DL}}x_{{k_2}}^{j,\rm{DL}}}}},&\text{if~}{{m \neq j}}.
	\end{array}} \right.
\end{split}
\end{equation}
\vspace{-1em}
\bibliographystyle{ieeetran}
\bibliography{Tactile}		
\bibliographystyle{ieeetr}
\end{document}